\documentclass[aps,prl,reprint,twocolumn,superscriptaddress]{revtex4-1}

\usepackage{graphicx}
\usepackage{epstopdf}
\usepackage{amssymb}
\usepackage{array}
\usepackage{amsmath}
\usepackage{bm}
\usepackage{color}
\usepackage{tabularx}
\usepackage[hidelinks]{hyperref}
\usepackage[normalem]{ulem}

\definecolor{blueblack}{rgb}{0, 0.2, 0.75}

\definecolor{lightgrey}{rgb}{0.8, 0.8, 0.8}

\begin{document}

%%%%%%%%%%%%%%%%%%%%%%%%%%%%%%%%%%%%%%%%%%%%%%%%%%%%%%%%%%%%
	
\title{Thermal Diffusivity Above the Mott-Ioffe-Regel Limit}

\author{Jiecheng Zhang}
\email[Corresponding author: ]{jiecheng@stanford.edu}
\affiliation{Department of Physics, Stanford University, Stanford, CA 94305,USA.}
\affiliation{Geballe Laboratory for Advanced Materials, Stanford University, Stanford, CA 94305, USA.}
\author{Erik D. Kountz}
\affiliation{Department of Physics, Stanford University, Stanford, CA 94305,USA.}
\affiliation{Geballe Laboratory for Advanced Materials, Stanford University, Stanford, CA 94305, USA.}
\author{Eli M. Levenson-Falk}
\affiliation{Department of Physics, University of Southern California, Los Angeles, CA 90089, USA.}
\author{Dongjoon Song}
\affiliation{Center for Correlated Electron Systems, Institute for Basic Science (IBS), Seoul 08826, Republic of Korea}
\author{Richard L. Greene}
\affiliation{Center for Nanophysics $\&$ Advanced Materials, University of Maryland, College Park, Maryland 20742, USA.}
\affiliation{Department of Physics, University of Maryland, College Park, Maryland 20742, USA.}
\author{Aharon Kapitulnik}
\affiliation{Department of Physics, Stanford University, Stanford, CA 94305,USA.}
\affiliation{Geballe Laboratory for Advanced Materials, Stanford University, Stanford, CA 94305, USA.}
\affiliation{Department of Applied Physics, Stanford University, Stanford, CA 94305, USA.}

\date{\today}

\begin{abstract}
We present high-resolution thermal diffusivity measurements on several near optimally doped electron- and hole-doped cuprate systems in a temperature range that passes through the Mott-Ioffe-Regel limit, above which the quasiparticle picture fails.  Our primary observations are that the inverse thermal diffusivity is linear in temperature and can be fitted to $D_Q^{-1}=aT+b$. The slope $a$ is interpreted through the Planckian relaxation time $\tau\approx\hbar/k_BT$ and a thermal diffusion velocity $v_B$, which is close, but larger than the sound velocity. The intercept $b$ represent a crossover diffusion constant that separates coherent from incoherent quasiparticles. These observations suggest that both phonons and electrons participate in the thermal transport, while reaching the Planckian limit for relaxation time.
\end{abstract}

\maketitle
%\section*{Introduction}
The standard paradigm for transport in Fermi-liquid metals relies on the existence of well-defined quasiparticles. Transport coefficients such as electrical and thermal conductivities can then be calculated using Boltzmann theory, where the electrons are treated semiclassically \cite{ziman1960}. However, such an approach fails when the quasiparticle mean free path becomes comparable to its de Broglie wavelengths. Beyond this so-called Mott-Ioffe-Regel (MIR) limit \cite{MIR}, the material is dubbed a ``bad metal'' \cite{EmeryKivelson1995},  and transport becomes ``incoherent,'' as the notion of momentum eigenstate quasiparticles fails.  A new theoretical framework is needed to describe this regime, while new experiments, complementary to the extensively studied electrical resistivity, are needed to provide an additional perspective on the problem. 

In the absence of a microscopic transport theory, one may still use thermodynamics supplemented by continuity equations for the charge and energy as conserved quantities. In the absence of thermoelectric effects, this approach leads to Einstein relations for the electrical conductivity $\sigma=\chi D_e$ and the thermal conductivity $\kappa=cD_Q$, which are expected to hold regardless of the presence of quasiparticles. Here, $\chi=e^2(dn/d\mu)$ is the charge susceptibility and is proportional to the density of states, $c$ is the specific heat, and $D_e$ and $D_Q$ are charge and thermal diffusivity, respectively. It is therefore interesting to explore transport in a non-quasiparticle regime by studying the diffusivities directly. Indeed, in recent studies diffusivity was singled out as a key observable for incoherent non-quasiparticle transport, possibly subject to fundamental quantum mechanical bounds \cite{Hartnoll2015}, particularly a minimum ``Planckian'' relaxation time $\tau\sim\hbar/k_BT$  \cite{Zaanen2004}, which leads to the iconic linear resistivity that persists beyond the MIR limit. Such a behavior has been observed in numerous resistivity measurements on strongly correlated materials including the hole-doped \cite{Takagi1992}  and electron-doped \cite{Bach2011} cuprate superconductors. However, no sufficient studies have been performed for thermal transport in these materials, despite evidence from early studies by Allen {\it et al.} \cite{Allen1994}, and more recent data from Zhang {\it et al.} \cite{Zhang5378}, suggesting that thermal transport in the high-temperature regime of the strongly correlated cuprate superconductors must involve incoherent electrons and phonons. 

In this letter we report new high-resolution thermal diffusivity measurements on single crystals of nearly optimally doped electron doped cuprates Sm$_{2-x}$Ce$_x$CuO$_4$ (SCCO), Nd$_{2-x}$Ce$_x$CuO$_4$ (NCCO), Pr$_{2-x}$Ce$_x$CuO$_4$ (PCCO), and near optimally doped Bi$_2$Sr$_2$CaCu$_2$O$_{8}$ (BSCCO) 
in the temperature range of 100 to 600K. These material exceed the MIR limit in resistivity \cite{MIR} above $\sim$250 to 300 K \cite{Bach2011}. Unlike the hole doped cuprates, the electron doped cuprates do not lose oxygen upon heating and thus can be studied to relatively high temperatures. In addition, comparing ``as grown'' with ``annealed'' samples allow for comparison of disorder effects on the high temperature thermal transport. Our primary observation is that for all samples we measured the inverse diffusivity is linear in temperature, and can be fitted to
\begin{equation}
	\label{theFit}
	D_Q^{-1}(T)= aT +b = \Big(  \alpha v_B^2 \frac{\hbar }{k_B T} \Big)^{-1}+D_0^{-1} \ ,
\end{equation}
The slope $a$ is interpreted as a result of the product of a diffusion velocity $v_B$, a Planckian relaxation time $\tau\approx\hbar/k_BT$, and an order-unity constant $\alpha$. The constant $b = D_0^{-1}$ represents a quantum-diffusion constant separating incoherent transport from a regime with well-defined quasiparticles, and will be discussed further below.  In the absence of an exact theoretical guidance we set $\alpha=1$, and estimate $v_B$ to be about twice to three times the sound velocity in that material. Comparison to literature data for diffusivity of undoped or lightly doped insulators of similar materials, we observe that  $b=0$ and the extracted velocity is close to the sound velocity of the respective material.  These observations unambiguously establish the substantial participation of electrons in the thermal transport and suggest that also the phonons reach the Planckian limit for relaxation time. 

Samples preparation  methods are discussed in the Supplemental Material. For the high resolution thermal diffusivity measurements we use a  photo-thermal microscope previously used in ref.~\cite{Zhang5378}, and described in details in the supplementary Materials section. Using this apparatus, the thermal diffusivity is obtained directly, without the need to measure the thermal conductivity and specific heat separately. An advantage of this apparatus, exploited in our previous study of underdoped YBCO, is the ability to measure the full in-plane anisotropy of the thermal diffusivity by orienting the pair of laser spots at any orientation with respect to the crystal axes. The mobility in the optics is further used for diagnostics of spatial uniformity of the thermal diffusivity values. Electrical resistivity was measured in the ab-plane on some of the samples for comparison.

The stability of oxygen in the electron-doped cuprates allow for measurements up to 600K. The ab-plane thermal diffusivities of two single crystal Sm$_{1.84}$Ce$_{0.16}$CuO$_4$, two Nd$_{1.85}$Ce$_{0.15}$CuO$_4$, an as grown and an annealed sample for each material, and a Pr$_{1.87}$Ce$_{0.13}$CuO$_4$ annealed sample were measured from 100-350K(SCCO as grown) and 100-600K(all others).  Thermal diffusivity of BSCCO was only measured to $\sim420$ K, above which the crystal loses oxygen in vacuum. Examples of our diffusivity results are shown in Fig.~\ref{AllDiff}, where the rest of the data is shown in the Supplemental Material.  These ranges of temperature capture the behavior below and above the resistive MIR limit. To appreciate the quality of the data, we refer to the Supplemental Material, which shows a typical frequency response of the phase shift from which we determine the value of $D_Q$. The single coefficient fit means that for time scales of $(20 {\rm kHz})^{-1}$ to $(200 {\rm Hz})^{-1}$ $D_Q$ is unique, indicating thermal transport of a uniform ``fluid''. We emphasize that all data presented in this paper show similar behavior. 
\begin{figure}[h]
	\centering
	\includegraphics[width=1.0\columnwidth]{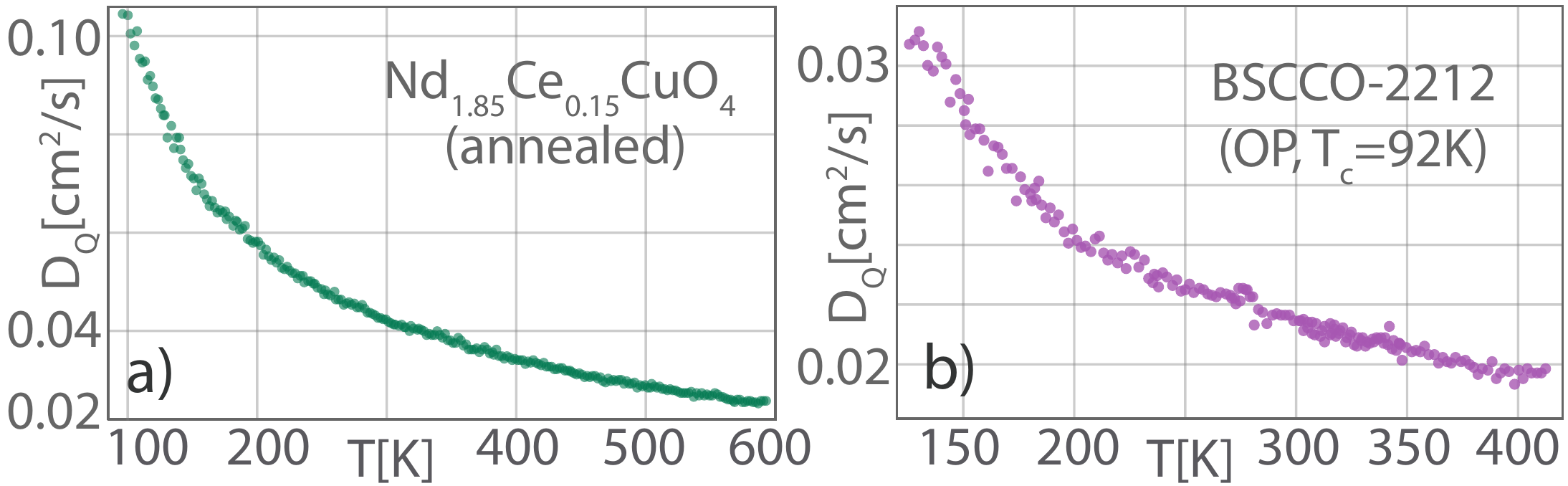}
	\caption{Examples for thermal diffusivity in the ab-plane measured as function of temperature using the optical setup. a)  Nd$_{1.85}$Ce$_{0.15}$CuO$_4$ annealed, and  b) Optimally doped Bi$_{2}$Sr$_{2}$CeCu$_2$O$_{8+x}$ with $T_c=92K$.  Statistical error is smaller than the data points. A systematic error of $\sim5\%$ is estimated as a result of  calibration of the optical paths, and the finite size of the focused laser spots.}
	\label{AllDiff}
\end{figure}

%%%%%%%%%%%%%%%%%%%%% FIRST NOTE MAGNITUDE
We first note that above $\sim 200$K, the thermal diffusivities of all these materials are very low\footnote{The thermal diffusivities we measured seem to be more natural if expressed in the alternative unit system, fathom$^2$/fortnight $\approx$ 0.03cm$^2$/cm.}, indicating that in addition to being bad electrical conductors, they are also bad thermal conductors, similar to complex insulators such as perovskites \cite{Hofmeister2010}.  An initial estimate of the electronic contribution to the thermal conductivity based on resistivity measurements and the Wiedemann-Franz law suggests that electrons contribute only about 10\% to 20\% of the total thermal transport in the relevant temperature range (see e.g.~\cite{Allen1994,Minami2003}). Thus, assuming thermal diffusivity that is dominated by phonons, we can estimate a thermal phonon mean-free-path $\ell_Q^\text{ph}$ from the measured thermal diffusivity $D_Q$ using the speed of sound $v_\text{s}$ as a typical phonon velocity: $D_Q\sim v_\text{s}\ell_Q^\text{ph}$. Following similar estimates for complex insulators  \cite{Hofmeister2010}, we use compressional sound velocity  \cite{Fil1996}, with the rational that much of the heat is transported by the longitudinal acoustic LA mode, since it involves excursions of atoms along the direction of heat propagation.  We estimate at room temperature $\ell_Q^\text{ph}(295K)\sim$6\AA~ for the electron doped cuprates, $\ell_Q^\text{ph}(295K)\sim$5\AA~ for near optimally doped Bi$_2$Sr$_2$CaCu$_2$O$_8$, and $\ell_Q^\text{ph}(295K)\sim$3\AA~ for underdoped YBa$_2$Cu$_3$O$_{6+x}$\cite{Zhang5378}. All these estimates yield mean free paths that are of order of the lattice constant ($\sim 4$\AA), which lead us to conclude that phonons are very strongly damped, with many modes poorly defined in this temperature range. Furthermore, the fact that the ``as grown'' and ``annealed'' samples show similar results suggest that disorder does not play a major role in the thermal transport in this high temperature regime. In fact, similar conclusion was already reached by Allen {\it et al.} \cite{Allen1994} in their study of thermal conductivity in the hole-doped Bi$_2$Sr$_2$CaCu$_2$O$_8$ and insulating Bi$_2$Sr$_2$YCu$_2$O$_8$ cuprates. Moreover, a recent comprehensive study of LA acoustic branch corresponding to the compressional sound velocity in BSCCO crystals ($v_s\approx 4.37 \times 10^5$ cm/s) show strong deviation from a simple harmonic mode, accompanied by interaction with optical modes already at 240K \cite{He2018}.  This conclusion is even more acute as we will argue below, that part of the measured thermal diffusivity is due to electrons, and in particular since the thermal diffusivity continue to decrease above room temperature according to Eqn.~\ref{theFit}. A more detailed comparison of the length scales is given in the Supplemental Material.

\begin{figure}[ht]
	\centering
 	\includegraphics[width=1.0\columnwidth]{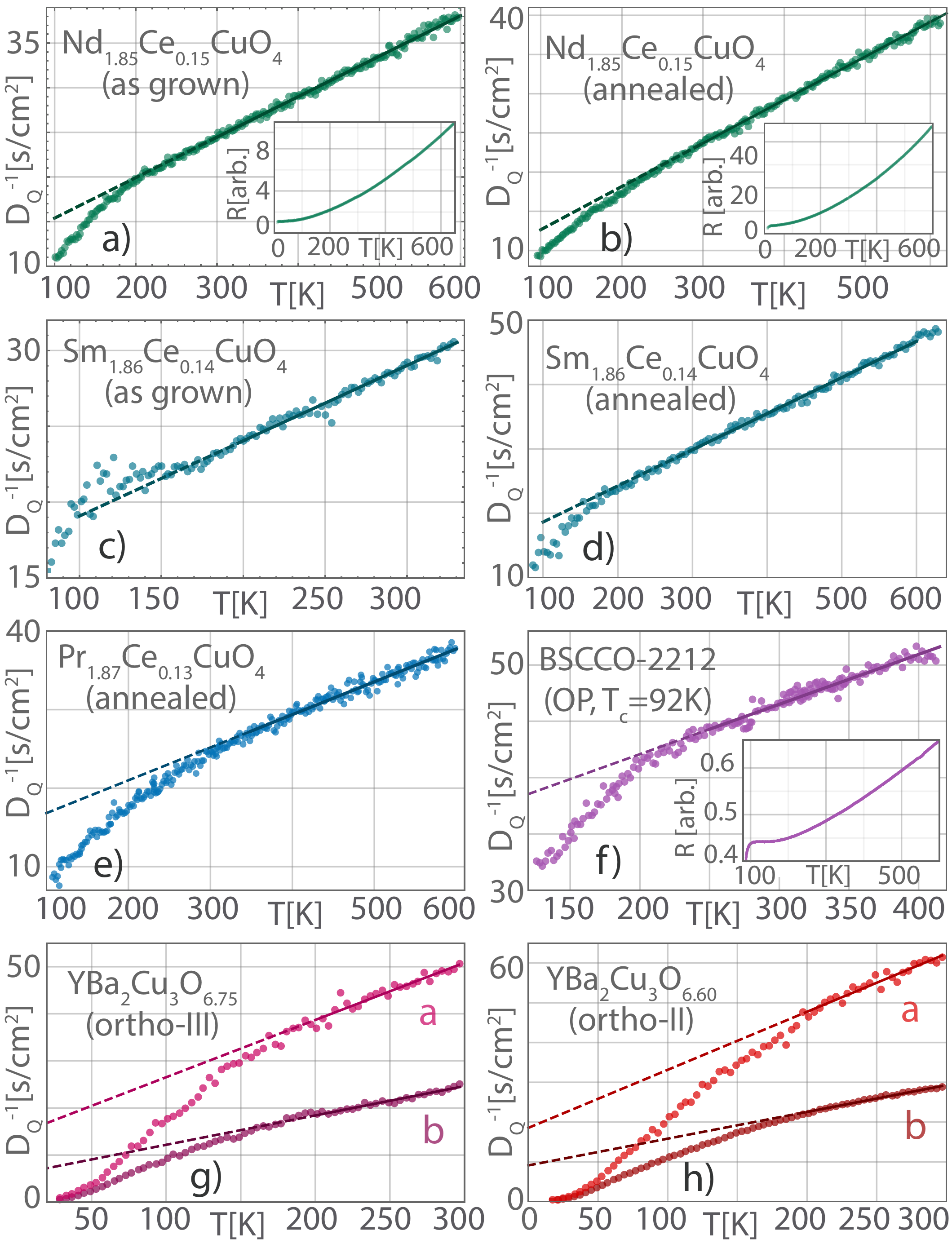}
 	\caption{Inverse thermal diffusivity $D_Q^{-1}(T)$ as function of temperature of optimal electron doped NCCO, SCCO, PCCO, optimal hole doped BSCCO crystal, and two underdoped YBCO crystals from \cite{Zhang5378}. Solid lines show linear fit in the form $D_Q^{-1}=aT+b$ to the data above 300K for electron doped cuprates, above 250K for BSCCO, and above 200K for YBCOs. Dotted lines show the same fit extended to lower temperatures. Insets in a) b) f) show 4-terminal resistance measured on the same crystals in arbitrary units.}
	\label{AllDiffInv}
\end{figure}

%%%%%%%%%%%%%%%%%%%%% SECOND NOTE T-LINEAR
Another notable feature of the data is that the inverse thermal diffusivities $D_Q^{-1}(T)$ at high temperatures are linear in $T$. Fig.~\ref{AllDiffInv} shows the same data of Fig.~\ref{AllDiff} plotted as inverse diffusivity, together with a high temperature fit to the form in Eqn.~\ref{theFit}.  For comparison, we also show resistance measurements on the NCCO and BSCCO samples in the same temperature range, where, as was previously established, they cross the MIR limit. These observations further support our previous assumption that underdoped YBCO crystals would have reached a similar linear dependence if they could be measured above room temperatures (the temperature range was limited to below room temperature to assure the same crystal structure and oxygen ordering and doping \cite{Liang2006}). Thus, for comparison we also show in Fig.~\ref{AllDiffInv} our previously published data on underdoped YBCO crystals \cite{Zhang5378}, together with a high temperature linear fit asymptote. Unique to this system was the similar behavior of the inverse diffusivity and the resistivity anisotropies, which was taken as a proof that both phonons and electrons participate in entropy transport above the MIR limit, possibly moving as a ``soup'' of overdamped electron-phonon fluid.

%%%%%%%%%%%%%%%%%%%%% DISCUSSION

% recap of phonon scattering mechanism
A $T^{-1}$ temperature dependence of the phonon relaxation time is ubiquitous in highly crystalline insulators (silicon, sapphire, etc.) at very high temperatures, traditionally explained as a consequence of Umklapp scattering of phonons with a scattering rate that decreases as $\theta_D/T$ \cite{Berman1976}, often commencing well below the Debye temperature \cite{Slack1979}. However, in the presence of impurities or other forms of disorder that limit the mean free path, as the temperature increases the thermal conductivity either saturates or weakly increases, approaching the pure crystals' high temperature limit (i.e. Umklapp limit) from below \cite{Slack1964,Cahill1992}. In most of these cases 
Umklapp scattering alone is unlikely to be sufficiently strong to account for the observed small mean-free path \cite{Vandersande1986,Langenberg2016}. Other relaxation channels, such as higher order phonon scattering and in particular scattering of acoustic phonons by interaction with optical phonons may dominate the thermal transport, particularly in complex material systems.  In the presence of electrons, phonon-electron scattering rate may be the dominating high-temperature cause of the observed short phonon mean free path (see e.g.  \cite{Steigmeier1964,Vandersande1986}).

% thermal conduction in complex insulators
Indeed, relevant to the present study are the complex insulators that host the electronic systems of the cuprates, where the unique crystal structure result in a very complex phonon-band structure, with optical phonon modes (many of them dispersive), that much exceed in energy the equivalent $k_BT$-range of the experiments. In addition, structurally complex systems such as perovskites typically exhibit tilting of octahedra, distortions of octahedra sites, disorder of the cation inside the octahedra, and when doped, anti-site disorder, all of which contribute to their low thermal conductivity \cite{Langenberg2016,Hofmeister2010}. A recent extensive study of insulating oxide perovskite compounds at elevated temperatures \cite{Hofmeister2010} revealed that their inverse thermal diffusivity exhibit a high temperature regime of $\ell^\text{ph}_Q<a$, that is dominated by a $T$-linear term. More recently, Martelli {\it et al.} \cite{Martelli2018} published detailed study of thermal transport in SrTiO$_3$, where at high temperatures a $D_Q^{-1}\propto T$ behavior was observed, and the prefactor was set by the sound velocity and Planckian relaxation time $\tau \approx \hbar/k_BT$ \cite{Phonons}.  The thermal conductivity at high temperatures can be further reduced by introducing low levels of Nb or Ca doping, pushing it closer to the minimum phonon thermal conductivity allowed in pure crystals. 

% linear fit D_Q(T)
The above discussion motivates us to interpret the observed $D_Q^{-1}(T)=aT+b$ behavior in a similar way to the complex insulators. Apart from the constant, ``$b$'', the thermal diffusion is characterized by a velocity $v_B$, a Planckian relaxation rate $\tau\approx\hbar/k_BT$ and an order-unity constant $\alpha$, which depends on the dimensionality of the system and the precise coefficient in the expression of $\tau$. The intercept, $D_0^{-1}$, which is the extrapolated $T\to 0$ limit of  $D_Q^{-1}(T)$, will be interpreted below as a quantum diffusion constant that separates the coherent from incoherent regimes of the electronic part of the thermal transport. This constant is zero for insulators (see e.g. \cite{Martelli2018}), which further demonstrate the fact that it is an incoherent current of  both,  electrons and phonons that transport entropy in the high temperature regime.

% analysis of the slope with \alpha=1
Since a prefactor for a Plankian relaxation time has not been rigorously calculated for these systems, choosing an appropriate $\alpha$ is somewhat arbitrary. A reasonable choice would be to take $\alpha=1$ (further discussion on the choice of $\alpha$ is given in the Supplementary Material).  The results for the diffusion velocity,  $v_B$ for the different materials are given in Table~\ref{fittable}. For comparison we also show the respective speed of sound $v_s$ and Fermi velocity $v_F$ for each of these material systems. Note that the fitted $v_B$ are very similar in magnitude, indicating very similar slope of $D_Q^{-1}$ in the T-linear regime.
\begin{table}[ht]
\begin{center}
	\caption{Effective thermal diffusion velocity $v_B$ and effective diffusion mass $m_D$ (in terms of free electron mass $m_e$) extracted from the fitting form Eq.~\ref{theFit} of measured thermal diffusivity. Here ``(n)'' denotes ``annealed'' and ``(g)'' denotes ``as grown''. For YBCO, ``(a)'' denotes $a$-axis and ``(b)'' denotes the chain $b$-axis. We also list the experimentally determined speed of sound $v_s$ \cite{Fil1996,Berggold2006} and Fermi velocities $v_F$\cite{Armitage2010,Vishik2010,Tallon1995}. }
	\label{fittable}
	\begin{tabular}{|c|c|c|c|c|c|c|}
		\hline\hline
				&$v_s$		&$v_B(\alpha=1)$	&$v_F$			&$m_D$		\\
		sample	&[$10^5$cm/s]	&[$10^6$cm/s]	&[$10^7$cm/s]	&[m$_e$]	\\
		\hline
		NCCO$_{0.15}$(g)	&$7.0$	&$1.7$	&$2.5$	&4.0	\\
		\hline
		NCCO$_{0.15}$(n)	&$7.0$	&$1.6$	&$2.5$	&2.6	\\ 
		\hline
		SCCO$_{0.16}$(g)	&$5.9$ 	&$1.7$	&$2.0$	&5.2	\\ 
		\hline
		SCCO$_{0.16}$(n)	&$5.9$ 	&$1.5$	&$2.0$	&5.0	\\ 		
		\hline
		PCCO$_{0.13}$(g)	&$6.25$	&$1.8$ &$2.1$	&4.6	\\ 	
		\hline
		BSCCO 				&$4.37$ 	&$1.7$	&$2.4$	&12		\\ 		
		\hline
		YBCO$_{6.75}$(a) 	&$6.05$ &$1.1$	&$2.25$	&5.3	\\ 	
		\hline
		YBCO$_{6.75}$(b) 	&$6.5$ 	&$1.5$	&$-$	&2.2	\\ 	
		\hline
		YBCO$_{6.60}$(a) 	&$6.05$ &$1.0$	&$2.1$	&6.8	\\ 	
		\hline
		YBCO$_{6.60}$(b) 	&$6.5$ 	&$1.4$	&$-$	&3.3	\\ 	
		\hline\hline
	\end{tabular}
	\end{center}
\end{table}

% diffusion mass m*
We now turn to the residual diffusion constant $D_0$. Since this $T\to 0$ intercept of the inverse diffusivity is observed only for the doped systems with appreciable carrier density, we conclude that this term is a consequence of the electronic contribution to the thermal diffusivity. However, since this term is extracted from above the MIR limit, where quasiparticles are not well defined, it will not make sense to interpret this term as a true zero-temperature limit obeying Matthiessen's rule. This empirical rule relies on coherent quasiparticles and its extrapolation to zero-temperature yields the impurity scattering. However, in the spirit of this rule, where the full scattering rate of quasiparticle is the sum of all scattering rates, we can interpret the constant $b=D_0^{-1}$ as a limiting rate for incoherent quasiparticles.  This then must be the quantum of diffusion at the MIR limit, where we substitute $k_F\ell=1$ into $D_{\text{MIR}}=\tfrac{1}{3}v_F\ell$, yielding $D_0=\tfrac{1}{3}\tfrac{\hbar}{m_D}$, where $m_D$ is an effective mass associated with the diffusion process. This term also describes the limiting rate of spread of a wavepacket of momentum-state quasiparticles of mass $m_D$ in time (see e.g. \cite{Merzbacher}), which should cease to be applied above the MIR limit. For example, effective mass extracted from quantum oscillations measurements on NCCO with Ce doping $x=0.15$ yielded $m^*\approx 3$m$_e$ \cite{Helm2015}, while the effective mass extracted from specific heat and Raman measurements on a similar Pr$_{2-x}$Ce$_{x}$CuO$_4$ (PCCO), with Ce $x=0.15$ is in the same range \cite{Balci2004,Qazilbash2005}. On the other hand, single particle mass enhancement extracted from optical conductivity data on PCCO and NCCO with similar doping, and at room temperature, yielded $m^*\approx 6$m$_e$  \cite{Millis2005}. For BSCCO2212 OP, $m^*$ is estimated from specific heat to be $8.4\pm 1.6m_e$\cite{Legros2019}, while for YBCO$_{6.60}$ optical conductivity\cite{Padilla2005} and quantum oscillation was \cite{Leyraud2015} found $m^* \sim 2-3 m_e$. While these $m^*$ values have been extracted in different temperature regimes, and using different techniques, they all seem to be quite similar to $m_D$, which further support our conjecture about $D_0$.

% resistivity
Finally we briefly discuss the resistivity above the MIR limit.  As we observe from 
the insets in Fig.~\ref{AllDiffInv}, as well as previous studies on similar material systems \cite{Bach2011,Scanderbeg2016}, the resistivity of the electron-doped cuprates behaves as $\rho\propto T^y$, with $y\geq 1$. The case of  $T^2$ resistivity has been discussed in detail recently \cite{Sarkar2018}, pointing out that a $T^2$ behavior at high temperatures is inconsistent with the standard Fermi liquid theory without hydrodynamic effects.  Assuming in the limit where all relaxation times are bounded by $\sim\hbar/k_BT$, thermal and charge diffusivities are expected to be equal, which, by using the Einstein relations, implies that the electronic susceptibility obeys
\begin{equation}
	\chi = (\rho D_e)^{-1} \approx (\rho D_Q)^{-1} \propto T^{1-y}
\end{equation}
While such estimation is in not an exact quantitative prediction, the decrease in $\chi$ as T increases is qualitatively consistent with the loss of carrier density at high temperature, which could be one interpretation of optical conductivity measurements in both hole-doped and electron-doped cuprates\cite{Takenaka2003,Homes2006}. Recent theoretical arguments for possible reduced electron susceptibility at high temperatures were discussed by Perepelitsky {\it et al.} \cite{Perepelitsky2016} in their high-temperature expansion studies of the Hubbard model, and by Werman {\it et al.} \cite{Werman2016}, for a specific model with strong electron-phonon interaction.

On the other hand, the near optimally-doped Bi$_2$Sr$_2$CeCu$_2$O$_{8}$ displays $T$-linear resistivity with typical slope found in many hole-doped cuprates (see e.g. \cite{ScatterSimilar}) , thus allow for a direct comparison of $D_Q$ and $D_e$. In particular we are interested in the velocity that controls the two diffusion constants above the MIR limit where the relaxation time reaches the Planckian limit. While for thermal diffusion we obtained a velocity $v_B\approx 1.7\times10^6$cm/s, the slope of the resistivity yields $v_e\approx v_F$, that is, the charge diffusivity seems to be controlled by the Fermi velocity as is also evident from the data of Bruin {\it et al.} \cite{ScatterSimilar}. Note that here we assume a constant charge susceptibility $\chi$. We believe that this result is particularly important for any attempt to construct a complete description of transport above the MIR limit as we show that the two processes, thermal and charge transports can be governed by the same Planckian relaxation time but different velocities.

\section*{Acknowledgments}

We thank Kamran Behnia, Sankar Das Sarma, Sean Hartnoll, Steve Kivelson, and Subir Sachdev for insightful discussions, to J. S. Higgins (U Maryland) for his technical help, and Y. He and K. Xu for their help in sample preparation. This work was supported by the Gordon and Betty Moore Foundation through Emergent Phenomena in Quantum Systems (EPiQS) Initiative Grant GBMF4529, and by the U. S. Department of Energy (DOE) Office of Basic Energy Science, Division of Materials Science and Engineering at Stanford under contract No. DE-AC02-76SF00515. Work at University of Maryland was supported by NSF grant DMR-1708334. 

\bibliography{SCCO}
\clearpage

%%%%%%%%%%%%%%%%%%%%%%%%%%%%%%%%%%%%%%%%%%%%%%%%%%%%%%%%%%
%%%%%%%%%%%%%%%%%%%%%%%%%%%%%%%%%%%%%%%%%%%%%%%%%%%%%%%%%%

\begin{center}
{\Large Supplementary Materials for}\\
{\bf Thermal Diffusivity Above the Mott-Ioffe-Regel Limit}\\
{\normalsize Jiecheng Zhang$^{1,2}$, Erik D. Kountz$^{1,2,3}$, Eli M. Levenson-Falk$^4$,\\Richard L. Greene$^{5,6}$, 
Aharon Kapitulnik$^{1,2,3,4,7\ast}$}\\
\end{center}

\bigskip
\section*{\large MATERIALS, METHODS AND ADDITIONAL INFORMATION}

%%%%%%%%%%%%%%%%%%%%%%%%%%%%%%%%%

\subsection*{Single Crystals Growth}

Sm$_{1.84}$Ce$_{0.16}$CuO$_4$ and Nd$_{1.85}$Ce$_{0.15}$CuO$_4$ single crystals were prepared by self-flux with a typical size of $0.5\times0.5\times0.03$ mm$^3$ as described in Ref.~\cite{Peng1991}. Some of the crystals were then reduced in a low-oxygen anneal to achieve a sharp superconducting transition. Bi$_2$Sr$_2$CaCu$_2$O$_{8+\delta}$ single crystals were grown using the traveling-solvent floating-zone technique describe in \cite{Eisaki2004}.

\subsection*{Thermal Diffusivity Measurements}

\subsubsection*{Principles of the Photothermal Apparatus}

\begin{figure}[ht]
	\centering
	\includegraphics[width=1.0\columnwidth]{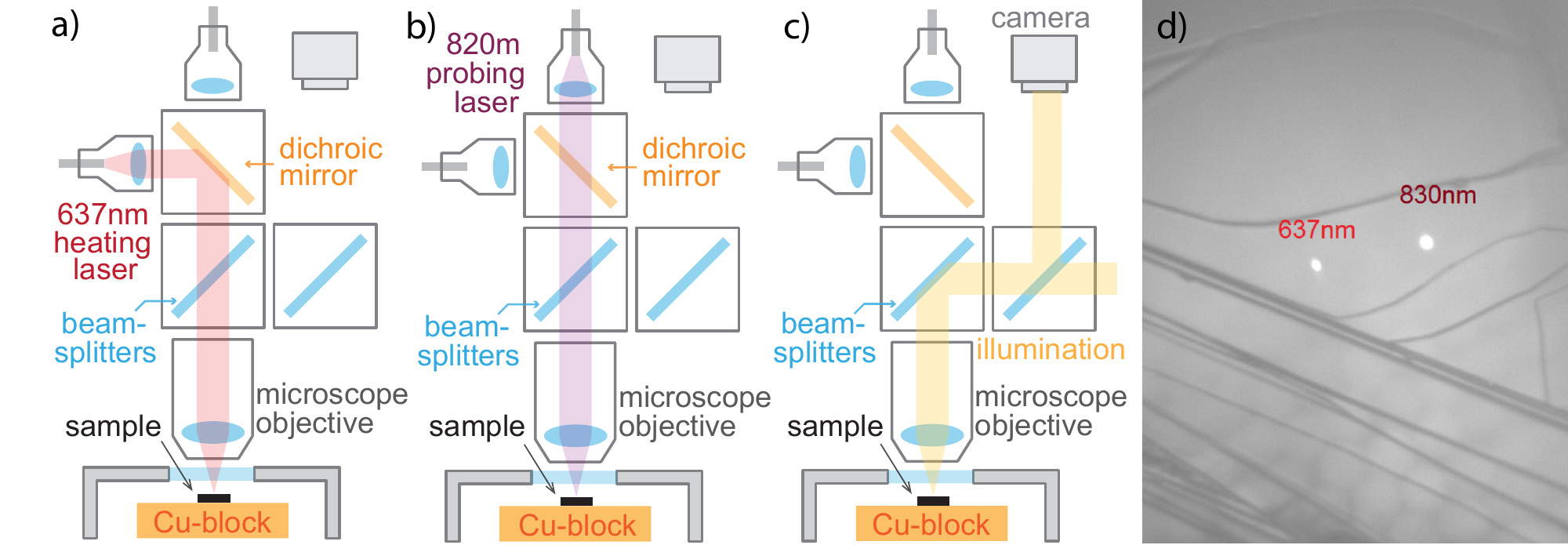}
	\caption{\label{setup}(color online) The schematic shows the optical paths of the setup. (a) Path of the heating laser (b) Path of the probing laser. The reflected light traverse the same path before gathered by a photodetector. (c) Path of illumination light source and camera vision. (d) Focused laser spots at typical measurement separation distance. The screenshot is taken on a TaS$_2$ sample surface for visual interest, where atomically flat terraces can be seen. The surface of crystals measured in the main text are mostly featureless. The dim hexagonal pattern visible in the lower left corner is the reflection on the fiber bundle used in the illumination source.}
\end{figure}

For the high resolution thermal diffusivity measurements we use a home-built photothermal microscope.  The microscope views the sample through a sapphire optical window in a cryostat, with the sample mounted to a cold finger just under the window.  A schematic is shown in Fig.~\ref{setup}. A heating laser at 637 nm and a probing laser at 820 nm are focused onto the sample surface by the microscope objective.  The focused spots have Gaussian radii of approximately 1$\mu$m and 2$\mu$m, respectively, due to the diffraction limit of different wavelengths, and can be moved independently over the sample surface.  A camera allows us to observe the sample surface nearby, align the spots in a particular orientation with respect to the crystal, and determine the distance between the spots.  
%Fig.~\ref{sample} shows a camera view of the two beams on different surfaces that we examined.

The output power of the heating laser is modulated with a sinusoidal profile $P(t)=P_0 [1 + \sin(\omega_0 t)]$. The modulation frequency $\omega_0/2\pi$ has a typical range of 500 Hz - 50 kHz, much slower than the microscopic equilibration on the order of picoseconds. This means that the parameters extracted are all within the in the DC limit of linear response, and that the dependency on the modulation frequency can be neglected. The probing laser is aimed at a spot a small distance (typically $10-20$ $\mu$m) away from the heating laser. The reflected light from the probing laser is diverted by an optical circulator and fed into a photodetector. The AC component of the photodetector signal is then fed to a lock-in amplifier referenced to the laser modulation and the amplitude and phase are measured. 

%%%%%%%%%%%%%%%%%%%%%%%%%%%%%%%%%

\subsubsection*{Measuring Thermal Diffusivity}
The diffusive transport of heat is governed by the diffusion equation
\begin{equation}
\label{diffusioneq}
	\frac{\partial\,\delta T(t,\vec{r})}{\partial t}
-D\nabla^2\,\delta T(t,\vec{r})
=\frac{q(t,\vec{r})}{c}
\end{equation}
where $\delta T$ is the temperature disturbance above the ambient temperature $T$, $\vec{r}=(x_1, x_2, x_3)$ is the spherical radial coordinate given in terms of the euclidean principal axes $x_i$, $q$ is the absorbed power density, $c$ is the volumetric specific heat capacity, $D \equiv \kappa / c$ is the thermal diffusivity, and $\kappa$ is the thermal conductivity.  Note that $c$ and $D$ are themselves functions of $T$, but in the limit of weak heating $\delta T << T$, we make the approximation $c(T+\delta T)\approx c(T)$ and $D(T+\delta T)\approx D(T)$. The temperature disturbance from both lasers is estimated to be $\lesssim$1 K through out the temperature range, so the previous approximation is valid.

The modulated power of the heating laser causes ripples in the temperature profile at the sample surface, the resulting change in reflectivity can then be picked up by the probing laser.  It is useful to write the response in frequency space $\tilde{\delta T}(\omega,\vec{r})$, where
\begin{equation}
\delta T(t,\vec{r}\,)=\int_{-\infty}^\infty\tilde{\delta T}(\omega,\vec{r})\exp(-i\omega t)d\omega 
\end{equation}
We model the focused heat source as a point source, $q(t,\vec{r}\,)= P_0 e^{-i\omega t}\delta^3(\vec{r})$.  This approximation is valid as long as the distance from the heating spot is much larger than the spot radius.  In a semi-infinite isotropic system, the temperature profile is spherically symmetric and takes the form
\begin{equation}
	\tilde{\delta T}(\omega,r)=\underbrace{\frac{P_0}{\kappa}
	\frac{1}{r}\exp \bigg(-\sqrt{\frac{\omega}{2D}}r\bigg)}_{\text{amplitude}}
		\underbrace{\exp\bigg(-i\sqrt{\frac{\omega}{2D}}r\bigg)}_{\text{phase}}
			\label{diffsol}
\end{equation}

\begin{figure}[ht]
	\centering
	\includegraphics[width=01.0\columnwidth]{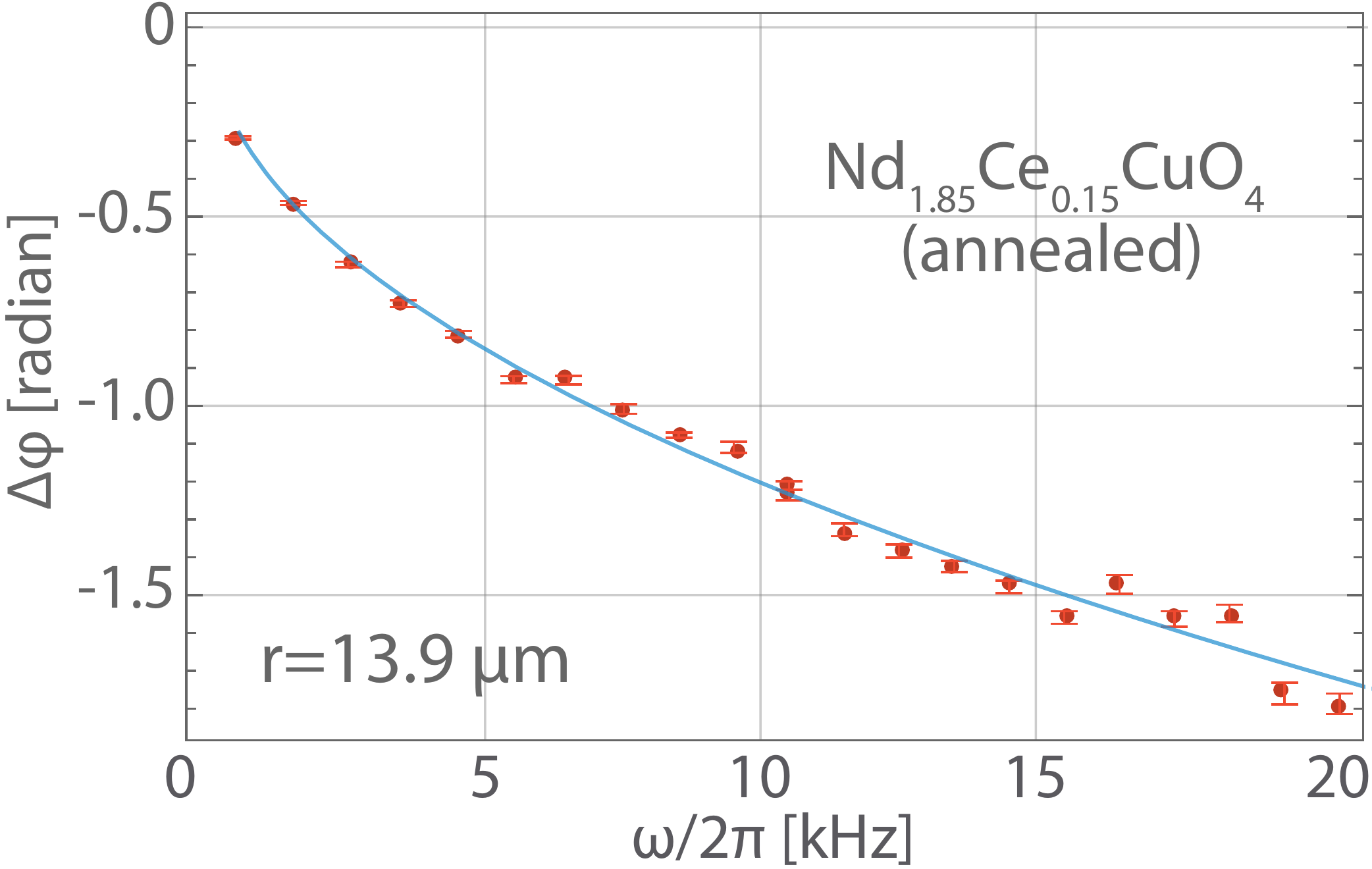}
	\caption{\label{phase}A typical phase delay as a function of heating laser modulation frequency, $\phi(\omega_0)$(red dot with orange error bars) obtained by sweeping $\omega_0$. This particular set is measured on an annealed $Nd_{1.85}Ce_{0.15}CuO_4$ at a separation distance of 13.9$\mu$m. The blue line is a fit using the form in Eqn.~\ref{diffsol}.}
\end{figure}

We vary the separation distance to verify the semi-infinite 3D system assumption and the small spot assumption, and we vary the heating power to verify the weak heating assumption. Our measurement gives us the response at the modulation frequency $\omega_0$. Because the separation distance between the lasers spots was measured using camera vision and each individually mounted sample comes at a small random tilt, a systematic error on the order of 5\% is associated to each set of measurements.  Although both the amplitude and the phase of the solution carry information about $D$, in actual measurements factors such as mechanical vibrations, fluctuations in the laser power, surface imperfections, and the temperature dependence of the differential reflectivity $dR/dT$ affect the amplitude of the reflectivity oscillation.  These factors only enter the solution by multiplying real numbers, and thus do not affect the phase of the signal, which is therefore the more robust probe. We obtain $D$ by fitting the phase delay $\phi$ between the source and the response signals as a function of $\omega$ at fixed $r$: $D = \omega r^2 / 2 \phi^2$. A typical fit is shown in Fig.~\ref{phase}. 
We check the homogeneity of the crystals by repeating measurements at different positions on the surface, and check the isotropy/anisotropy by rotating the relative orientation between the laser spots. 

\begin{figure}[h]
	\centering
	\includegraphics[width=1.0\columnwidth]{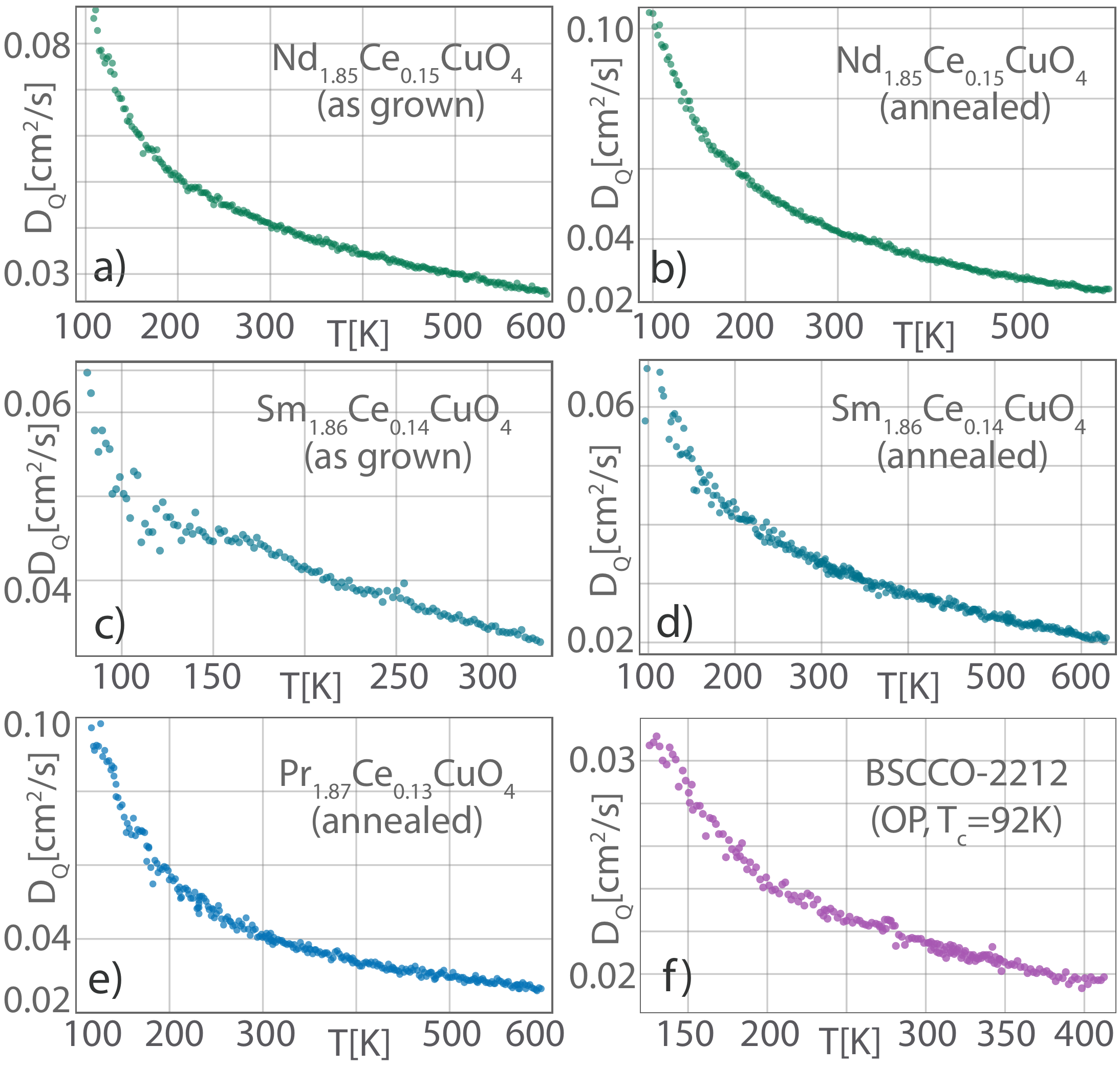}
	\caption{Thermal diffusivity in the ab-plane measured as function of temperature using the optical setup. a) Nd$_{1.85}$Ce$_{0.15}$CuO$_4$ as grown, 100K-600K. b) Nd$_{1.85}$Ce$_{0.15}$CuO$_4$ annealed, 100K-600K c) Sm$_{1.84}$Ce$_{0.16}$CuO$_4$ as grown, 100K-350K. (The irregularity around 120K is likely due to disturbance in the optics instead of intrinsic properties of the material). d) Sm$_{1.84}$Ce$_{0.16}$CuO$_4$ annealed, 100K-600K. e) Pr$_{1.87}$Ce$_{0.13}$CuO$_4$ as grown, 100K-600K. f) Optimally doped Bi$_{2}$Sr$_{2}$CeCu$_2$O$_{8+x}$ with $T_c=92K$, 120K-415K.}
	\label{AllDiffs}
\end{figure}

The results of the diffusivity measurements of all the samples discussed in the main text are shown in Fig.~\ref{AllDiffs}.

%%%%%%%%%%%%%%%%%%%%%%%%%%%%%%%%%

\subsection*{Resistivity Measurements}

The ab-plane resistances of the  samples were measured in a van der Pauw configuration. On the electron doped cuprates, four contacts were made on each of the corners of a sample by depositing 200nm of gold, annealing for 1 hour at 500$^\circ$C, applying DuPont silver paste over the gold patches, and annealing again for 0.5 hour at 550$^\circ$C. For the Bi-2212 sample, contacts were made by depositing 200nm of gold without any further annealing. All final contact resistances are on the order of 2$\Omega$.
Since we are only interested in the temperature dependence of the resistance, the measured values were not converted to resistivity.

%%%%%%%%%%%%%%%%%%%%%%%%%%%%%%%%%

\subsection*{Estimated Phonon Thermal Mean Free Path}

Within a simple kinetic theory, both electrons and phonons contribute to the thermal transport such that the total thermal conductivity is
\[\kappa= cD_Q= \kappa_\text{ph}+\kappa_\text{e}=c_\text{ph}D_\text{ph}+c_\text{e}D_\text{e} \]
where the total specific heat is $c=c_e+c_\text{ph}$. At room temperature $c_\text{ph}\approx c$, while $c_e\ll c_\text{ph}$. 
Thus, if we assume no electronic contribution to thermal transport, we approximate
\[  D_\text{ph}=\frac{c}{c_\text{ph}}D_Q - \frac{c_\text{e}}{c_\text{ph}}D_\text{e} \approx D_Q \]
The mean free path is then estimated from the relation $D_Q\sim v_\text{s}\ell_Q^\text{ph}$, where $v_s$ is the sound velocity. This approach is expected to fail when $\ell_Q^\text{ph}\approx a$, where $a$ is the lattice constant ($a \approx 4$\AA~ for all materials studied in this paper). We summarize in Table~\ref{tt} the room temperature values of the materials parameters, and the calculated phonon mean free path with the assumption that all the diffusivity is due to phonons.
\begin{table*}[ht]
\begin{tabular}{|c|c|c|c|c|c|c|c|c|c|}
\hline
Sample& $D_Q$ & c &$\kappa^*=cD_Q$ &$\kappa$ &  $v_s$ & $v_F$ & ~$\ell_{Q}^\text{ph}\approx D_Q/v_s$  &$\kappa_e(L_0)$ & $\theta_D$ \\
&cm$^2$/s & ~J/cm$^3$-K & W/cm-K &W/cm-K &  ~$\times10^5$cm/s & ~$\times10^7$cm/s &\AA&  W/cm-K  & K \\
  \hline
NCCO$_{0.15}$(g)&  0.041  & 2.89  &0.056	& N/A  & 7.0   &2.5 & 5.8    & 0.009 &  420 \\ 
NCCO$_{0.15}$(n)&  0.041  & 2.89  &0.12	& 0.092  & 7.0 &2.5   & 5.8    & 0.02 &  420 \\ 
SCCO$_{0.16}$(g)  &  0.034  & 2.91  &0.099& N/A  & 5.9  &2.0  & 5.7    & 0.009 &  320\\ 
SCCO$_{0.16}$(n) &  0.032 & 2.91 &0.093	& N/A & 5.9  &2.0 & 5.4   & 0.025 &  320\\ 
PCCO$_{0.16}$(n) &  0.042  & 2.58  &0.108	& 0.080  & 6.25 &2.1   & 6.7    & 0.015 &  360\\ 
BSCCO:2212   &  0.021   & 2.35&0.05	& 0.058   &4.37 &2.4 & 4.8   & 0.009 &  280\\ 
YBCO$_{6.6}$($a$-dir)   & 0.017   & 2.7  &0.046	& 0.05-0.065   &6.05 &2.1  & 3.1  & 0.013 &  410\\ 
YBCO$_{6.6}$($b$-dir)   & 0.034    & 2.7  &0.092	& 0.05-0.065   &6.5  & N/A  & 6.2 & 0.013 &  410\\ 
YBCO$_{6.75}$($a$-dir)   & 0.019    & 2.7 & 0.051	& 0.047-0.068  & 6.05 &2.25  & 2.8 & 0.017&  410 \\ 
YBCO$_{6.75}$($b$-dir)   & 0.038    & 2.7  &0.092	& 0.047-0.068    &6.5  & N/A  & 5.6  & 0.013 &  410\\ 
\hline
\end{tabular}
\caption{Room temperature thermal transport parameters for cuprates discussed in the paper. $D_Q$ is the measured diffusivity while the specific heat $c$, thermal conductivity $\kappa$, sound velocity $v_s$ and Fermi velocity $v_F$ are literature values referenced in the text. $\kappa^*$ and $\ell_{Q}^{ph}$ are calculated using measured $D_Q$ and literature data. BSCCO data is for optimally doped Bi$_2$Sr$_2$CaCu$_2$O$_8$ as found in the literature. YBCO$_{6.60}$($a$-dir) and YBCO$_{6.75}$($a$-dir)  $a$-direction data is on single crystals from similar doping, while diffusivity data is from the present work.  $\kappa_e$ represent an expecxted contribution to the thermal conductivity based on Wiedemann-Franz law, using $L_0=2.44\times 10^{-8}$ W$\Omega$/K$^2$ as the Lorentz number.}
  \label{tt}
\end{table*}

%%%%%%%%%%%%%%%%%%%%%%%%%%%%%%%%%

\subsection*{Alternative choice of $\alpha$ from the insulators}
In the main text we have used $\alpha=1$ to extract $v_B$ from the slope of the high temperature T-linear region in inverse thermal diffusivity:
\[ D_Q^{-1}=(\alpha\,v_B^2 \frac{\hbar}{k_B})^{-1}\,T+D_0^{-1} \]
The coefficient $\alpha$ is an unknown, order one constant in the definition of the Plankian relaxation time, $\tau\approx \hbar/(k_BT)$, and may also include the dimensionality of the diffusion process. It is possible that the precise value of this coefficient is sensitive to the microscopic details of system, unlike the linear dependence itself which is expected to be a universal feature. As we argued in the text, heat transport is dominated by the lattice,. with electrons adding to the overall diffusion process. Thus, it may make sense to extract $\alpha$ from the equivalent parent insulating, or lightly doped insulators of the same itinerant systems. In fact, many complex insulators also exhibit a high-temperature T-linear of $D_Q^{-1}$\cite{Langenberg2016,Hofmeister2010,Martelli2018}, with an extracted mean free path that approaches, or even becomes a fraction of the lattice constant. Careful study of the inverse diffusivity also reveal that the extrapolated $T\to 0$ inverse diffusivity vanishes. Since for insulators the relevant velocity is the sound velocity, for these systems the determination of $\alpha$ is unique.

Relevant to our data, Nd$_2$CuO$_4$ is the parent compound of the doped NCCO studied here. However, strong magnon contribution to the thermal conductivity may be a dominant effect. However, with light doping such as in Nd$_{1.975}$Ce$_{0.025}$CuO$_4$, magnetism is  weakened while the material is still an insulator, and much of the heat is transported by phonons. Using existing thermal conductivity\cite{Cohn1992} and specific heat\cite{Denisova2014} measurements of Nd$_{1.975}$Ce$_{0.025}$CuO$_4$, we estimated in this insulating compound $\alpha\sim 2.0$, which in turn gives the value of entropy diffusion velocity as $v_B\sim 1.13\times 10^{6}$cm/s. This estimated value is still larger than $v_s\approx 7.0\times 10^5$cm/s in NCCO, which together with the finite intercept $D_0^{-1}$ strengthen our claims that electrons make a substantial contribution to the thermal transport in these materials.

\bibliography{SCCO}

\begin{thebibliography}{47}%
\makeatletter
\providecommand \@ifxundefined [1]{%
 \@ifx{#1\undefined}
}%
\providecommand \@ifnum [1]{%
 \ifnum #1\expandafter \@firstoftwo
 \else \expandafter \@secondoftwo
 \fi
}%
\providecommand \@ifx [1]{%
 \ifx #1\expandafter \@firstoftwo
 \else \expandafter \@secondoftwo
 \fi
}%
\providecommand \natexlab [1]{#1}%
\providecommand \enquote  [1]{``#1''}%
\providecommand \bibnamefont  [1]{#1}%
\providecommand \bibfnamefont [1]{#1}%
\providecommand \citenamefont [1]{#1}%
\providecommand \href@noop [0]{\@secondoftwo}%
\providecommand \href [0]{\begingroup \@sanitize@url \@href}%
\providecommand \@href[1]{\@@startlink{#1}\@@href}%
\providecommand \@@href[1]{\endgroup#1\@@endlink}%
\providecommand \@sanitize@url [0]{\catcode `\\12\catcode `\$12\catcode
  `\&12\catcode `\#12\catcode `\^12\catcode `\_12\catcode `\%12\relax}%
\providecommand \@@startlink[1]{}%
\providecommand \@@endlink[0]{}%
\providecommand \url  [0]{\begingroup\@sanitize@url \@url }%
\providecommand \@url [1]{\endgroup\@href {#1}{\urlprefix }}%
\providecommand \urlprefix  [0]{URL }%
\providecommand \Eprint [0]{\href }%
\providecommand \doibase [0]{http://dx.doi.org/}%
\providecommand \selectlanguage [0]{\@gobble}%
\providecommand \bibinfo  [0]{\@secondoftwo}%
\providecommand \bibfield  [0]{\@secondoftwo}%
\providecommand \translation [1]{[#1]}%
\providecommand \BibitemOpen [0]{}%
\providecommand \bibitemStop [0]{}%
\providecommand \bibitemNoStop [0]{.\EOS\space}%
\providecommand \EOS [0]{\spacefactor3000\relax}%
\providecommand \BibitemShut  [1]{\csname bibitem#1\endcsname}%
\let\auto@bib@innerbib\@empty
%</preamble>
\bibitem [{\citenamefont {Ziman}(1960)}]{ziman1960}%
  \BibitemOpen
  \bibfield  {author} {\bibinfo {author} {\bibfnamefont {J.~M.}\ \bibnamefont
  {Ziman}},\ }\href@noop {} {\emph {\bibinfo {title} {Electrons and Phonons:
  The Theory of Transport Phenomena in Solids}}}\ (\bibinfo  {publisher}
  {Oxford University Press},\ \bibinfo {address} {Oxford, UK},\ \bibinfo {year}
  {1960})\BibitemShut {NoStop}%
\bibitem [{MIR()}]{MIR}%
  \BibitemOpen
  \href@noop {} {}\bibinfo {note} {The Mott-Ioffe-Regel limit has been
  expressed in different ways in the literature, e.g. as $k_F\ell \approx 1$,
  where $k_F$ is the Fermi wavevector and $\ell$ is the mean free path, or
  $\ell/a\approx 1$ where $a$ is the lattice constant. These approaches
  typically produce the same order of magnitude estimate. In this paper we use
  the criterion proposed by Emery and Kivelson in \cite{EmeryKivelson1995} of
  $\ell/\lambda_F \approx 1$ where $\lambda_F = 2\pi/k_F$.}\BibitemShut {Stop}%
\bibitem [{\citenamefont {Emery}\ and\ \citenamefont
  {Kivelson}(1995)}]{EmeryKivelson1995}%
  \BibitemOpen
  \bibfield  {author} {\bibinfo {author} {\bibfnamefont {V.~J.}\ \bibnamefont
  {Emery}}\ and\ \bibinfo {author} {\bibfnamefont {S.~A.}\ \bibnamefont
  {Kivelson}},\ }\href {\doibase 10.1103/PhysRevLett.74.3253} {\bibfield
  {journal} {\bibinfo  {journal} {Phys. Rev. Lett.}\ }\textbf {\bibinfo
  {volume} {74}},\ \bibinfo {pages} {3253} (\bibinfo {year}
  {1995})}\BibitemShut {NoStop}%
\bibitem [{\citenamefont {Hartnoll}(2015)}]{Hartnoll2015}%
  \BibitemOpen
  \bibfield  {author} {\bibinfo {author} {\bibfnamefont {S.~A.}\ \bibnamefont
  {Hartnoll}},\ }\href {\doibase 10.1038/NPHYS3174} {\bibfield  {journal}
  {\bibinfo  {journal} {Nature Physics}\ }\textbf {\bibinfo {volume} {11}},\
  \bibinfo {pages} {54} (\bibinfo {year} {2015})}\BibitemShut {NoStop}%
\bibitem [{\citenamefont {Zaanen}(2004)}]{Zaanen2004}%
  \BibitemOpen
  \bibfield  {author} {\bibinfo {author} {\bibfnamefont {J.}~\bibnamefont
  {Zaanen}},\ }\href@noop {} {\bibfield  {journal} {\bibinfo  {journal}
  {Nature}\ }\textbf {\bibinfo {volume} {430}},\ \bibinfo {pages} {512 EP}
  (\bibinfo {year} {2004})}\BibitemShut {NoStop}%
\bibitem [{\citenamefont {Takagi}\ \emph {et~al.}(1992)\citenamefont {Takagi},
  \citenamefont {Batlogg}, \citenamefont {Kao}, \citenamefont {Kwo},
  \citenamefont {Cava}, \citenamefont {Krajewski},\ and\ \citenamefont
  {Peck}}]{Takagi1992}%
  \BibitemOpen
  \bibfield  {author} {\bibinfo {author} {\bibfnamefont {H.}~\bibnamefont
  {Takagi}}, \bibinfo {author} {\bibfnamefont {B.}~\bibnamefont {Batlogg}},
  \bibinfo {author} {\bibfnamefont {H.~L.}\ \bibnamefont {Kao}}, \bibinfo
  {author} {\bibfnamefont {J.}~\bibnamefont {Kwo}}, \bibinfo {author}
  {\bibfnamefont {R.~J.}\ \bibnamefont {Cava}}, \bibinfo {author}
  {\bibfnamefont {J.~J.}\ \bibnamefont {Krajewski}}, \ and\ \bibinfo {author}
  {\bibfnamefont {W.~F.}\ \bibnamefont {Peck}},\ }\href {\doibase
  10.1103/PhysRevLett.69.2975} {\bibfield  {journal} {\bibinfo  {journal}
  {Phys. Rev. Lett.}\ }\textbf {\bibinfo {volume} {69}},\ \bibinfo {pages}
  {2975} (\bibinfo {year} {1992})}\BibitemShut {NoStop}%
\bibitem [{\citenamefont {Bach}\ \emph {et~al.}(2011)\citenamefont {Bach},
  \citenamefont {Saha}, \citenamefont {Kirshenbaum}, \citenamefont {Paglione},\
  and\ \citenamefont {Greene}}]{Bach2011}%
  \BibitemOpen
  \bibfield  {author} {\bibinfo {author} {\bibfnamefont {P.~L.}\ \bibnamefont
  {Bach}}, \bibinfo {author} {\bibfnamefont {S.~R.}\ \bibnamefont {Saha}},
  \bibinfo {author} {\bibfnamefont {K.}~\bibnamefont {Kirshenbaum}}, \bibinfo
  {author} {\bibfnamefont {J.}~\bibnamefont {Paglione}}, \ and\ \bibinfo
  {author} {\bibfnamefont {R.~L.}\ \bibnamefont {Greene}},\ }\href {\doibase
  10.1103/PhysRevB.83.212506} {\bibfield  {journal} {\bibinfo  {journal} {Phys.
  Rev. B}\ }\textbf {\bibinfo {volume} {83}},\ \bibinfo {pages} {212506}
  (\bibinfo {year} {2011})}\BibitemShut {NoStop}%
\bibitem [{\citenamefont {Allen}\ \emph {et~al.}(1994)\citenamefont {Allen},
  \citenamefont {Du}, \citenamefont {Mihaly},\ and\ \citenamefont
  {Forro}}]{Allen1994}%
  \BibitemOpen
  \bibfield  {author} {\bibinfo {author} {\bibfnamefont {P.~B.}\ \bibnamefont
  {Allen}}, \bibinfo {author} {\bibfnamefont {X.~Q.}\ \bibnamefont {Du}},
  \bibinfo {author} {\bibfnamefont {L.}~\bibnamefont {Mihaly}}, \ and\ \bibinfo
  {author} {\bibfnamefont {L.}~\bibnamefont {Forro}},\ }\href {\doibase
  10.1103/PhysRevB.49.9073} {\bibfield  {journal} {\bibinfo  {journal} {Phys.
  Rev. B}\ }\textbf {\bibinfo {volume} {49}},\ \bibinfo {pages} {9073}
  (\bibinfo {year} {1994})}\BibitemShut {NoStop}%
\bibitem [{\citenamefont {Zhang}\ \emph {et~al.}(2017)\citenamefont {Zhang},
  \citenamefont {Levenson-Falk}, \citenamefont {Ramshaw}, \citenamefont {Bonn},
  \citenamefont {Liang}, \citenamefont {Hardy}, \citenamefont {Hartnoll},\ and\
  \citenamefont {Kapitulnik}}]{Zhang5378}%
  \BibitemOpen
  \bibfield  {author} {\bibinfo {author} {\bibfnamefont {J.}~\bibnamefont
  {Zhang}}, \bibinfo {author} {\bibfnamefont {E.~M.}\ \bibnamefont
  {Levenson-Falk}}, \bibinfo {author} {\bibfnamefont {B.~J.}\ \bibnamefont
  {Ramshaw}}, \bibinfo {author} {\bibfnamefont {D.~A.}\ \bibnamefont {Bonn}},
  \bibinfo {author} {\bibfnamefont {R.}~\bibnamefont {Liang}}, \bibinfo
  {author} {\bibfnamefont {W.~N.}\ \bibnamefont {Hardy}}, \bibinfo {author}
  {\bibfnamefont {S.~A.}\ \bibnamefont {Hartnoll}}, \ and\ \bibinfo {author}
  {\bibfnamefont {A.}~\bibnamefont {Kapitulnik}},\ }\href {\doibase
  10.1073/pnas.1703416114} {\bibfield  {journal} {\bibinfo  {journal}
  {Proceedings of the National Academy of Sciences}\ }\textbf {\bibinfo
  {volume} {114}},\ \bibinfo {pages} {5378 } (\bibinfo {year}
  {2017})}\BibitemShut {NoStop}%
\bibitem [{Note1()}]{Note1}%
  \BibitemOpen
  \bibinfo {note} {The thermal diffusivities we measured seem to be more
  natural if expressed in the alternative unit system, fathom$^2$/fortnight
  $\approx $ 0.03cm$^2$/cm.}\BibitemShut {Stop}%
\bibitem [{\citenamefont {Hofmeister}(2010)}]{Hofmeister2010}%
  \BibitemOpen
  \bibfield  {author} {\bibinfo {author} {\bibfnamefont {A.~M.}\ \bibnamefont
  {Hofmeister}},\ }\href {\doibase 10.1063/1.3371815} {\bibfield  {journal}
  {\bibinfo  {journal} {Journal of Applied Physics}\ }\textbf {\bibinfo
  {volume} {107}},\ \bibinfo {pages} {103532} (\bibinfo {year}
  {2010})}\BibitemShut {NoStop}%
\bibitem [{\citenamefont {Minami}\ \emph {et~al.}(2003)\citenamefont {Minami},
  \citenamefont {Wittorff}, \citenamefont {Yelland}, \citenamefont {Cooper},
  \citenamefont {Changkang},\ and\ \citenamefont {Hodby}}]{Minami2003}%
  \BibitemOpen
  \bibfield  {author} {\bibinfo {author} {\bibfnamefont {H.}~\bibnamefont
  {Minami}}, \bibinfo {author} {\bibfnamefont {V.~W.}\ \bibnamefont
  {Wittorff}}, \bibinfo {author} {\bibfnamefont {E.~A.}\ \bibnamefont
  {Yelland}}, \bibinfo {author} {\bibfnamefont {J.~R.}\ \bibnamefont {Cooper}},
  \bibinfo {author} {\bibfnamefont {C.}~\bibnamefont {Changkang}}, \ and\
  \bibinfo {author} {\bibfnamefont {J.~W.}\ \bibnamefont {Hodby}},\ }\href
  {\doibase 10.1103/PhysRevB.68.220503} {\bibfield  {journal} {\bibinfo
  {journal} {Phys. Rev. B}\ }\textbf {\bibinfo {volume} {68}},\ \bibinfo
  {pages} {220503} (\bibinfo {year} {2003})}\BibitemShut {NoStop}%
\bibitem [{\citenamefont {Fil}\ \emph {et~al.}(1996)\citenamefont {Fil},
  \citenamefont {Kolobov}, \citenamefont {Fil}, \citenamefont {Barilo},\ and\
  \citenamefont {Zhigunov}}]{Fil1996}%
  \BibitemOpen
  \bibfield  {author} {\bibinfo {author} {\bibfnamefont {D.}~\bibnamefont
  {Fil}}, \bibinfo {author} {\bibfnamefont {I.}~\bibnamefont {Kolobov}},
  \bibinfo {author} {\bibfnamefont {V.}~\bibnamefont {Fil}}, \bibinfo {author}
  {\bibfnamefont {S.}~\bibnamefont {Barilo}}, \ and\ \bibinfo {author}
  {\bibfnamefont {D.}~\bibnamefont {Zhigunov}},\ }\href {\doibase
  10.1007/BF02571069} {\bibfield  {journal} {\bibinfo  {journal} {Czechoslovak
  Journal of Physics}\ }\textbf {\bibinfo {volume} {46}},\ \bibinfo {pages}
  {2155} (\bibinfo {year} {1996})},\ \bibinfo {note} {21st International
  Conference on Low Temperature Physics (LT 21), Prague, Czech Republic, AUG
  08-14, 1996}\BibitemShut {NoStop}%
\bibitem [{\citenamefont {He}\ \emph {et~al.}(2018)\citenamefont {He},
  \citenamefont {Wu}, \citenamefont {Song}, \citenamefont {Lee}, \citenamefont
  {Said}, \citenamefont {Alatas}, \citenamefont {Bosak}, \citenamefont
  {Girard}, \citenamefont {Souliou}, \citenamefont {Ruiz}, \citenamefont
  {Hepting}, \citenamefont {Bluschke}, \citenamefont {Schierle}, \citenamefont
  {Weschke}, \citenamefont {Lee}, \citenamefont {Jang}, \citenamefont {Huang},
  \citenamefont {Hashimoto}, \citenamefont {Lu}, \citenamefont {Song},
  \citenamefont {Yoshida}, \citenamefont {Eisaki}, \citenamefont {Shen},
  \citenamefont {Birgeneau}, \citenamefont {Yi},\ and\ \citenamefont
  {Frano}}]{He2018}%
  \BibitemOpen
  \bibfield  {author} {\bibinfo {author} {\bibfnamefont {Y.}~\bibnamefont
  {He}}, \bibinfo {author} {\bibfnamefont {S.}~\bibnamefont {Wu}}, \bibinfo
  {author} {\bibfnamefont {Y.}~\bibnamefont {Song}}, \bibinfo {author}
  {\bibfnamefont {W.-S.}\ \bibnamefont {Lee}}, \bibinfo {author} {\bibfnamefont
  {A.~H.}\ \bibnamefont {Said}}, \bibinfo {author} {\bibfnamefont
  {A.}~\bibnamefont {Alatas}}, \bibinfo {author} {\bibfnamefont
  {A.}~\bibnamefont {Bosak}}, \bibinfo {author} {\bibfnamefont
  {A.}~\bibnamefont {Girard}}, \bibinfo {author} {\bibfnamefont {S.~M.}\
  \bibnamefont {Souliou}}, \bibinfo {author} {\bibfnamefont {A.}~\bibnamefont
  {Ruiz}}, \bibinfo {author} {\bibfnamefont {M.}~\bibnamefont {Hepting}},
  \bibinfo {author} {\bibfnamefont {M.}~\bibnamefont {Bluschke}}, \bibinfo
  {author} {\bibfnamefont {E.}~\bibnamefont {Schierle}}, \bibinfo {author}
  {\bibfnamefont {E.}~\bibnamefont {Weschke}}, \bibinfo {author} {\bibfnamefont
  {J.-S.}\ \bibnamefont {Lee}}, \bibinfo {author} {\bibfnamefont
  {H.}~\bibnamefont {Jang}}, \bibinfo {author} {\bibfnamefont {H.}~\bibnamefont
  {Huang}}, \bibinfo {author} {\bibfnamefont {M.}~\bibnamefont {Hashimoto}},
  \bibinfo {author} {\bibfnamefont {D.-H.}\ \bibnamefont {Lu}}, \bibinfo
  {author} {\bibfnamefont {D.}~\bibnamefont {Song}}, \bibinfo {author}
  {\bibfnamefont {Y.}~\bibnamefont {Yoshida}}, \bibinfo {author} {\bibfnamefont
  {H.}~\bibnamefont {Eisaki}}, \bibinfo {author} {\bibfnamefont {Z.-X.}\
  \bibnamefont {Shen}}, \bibinfo {author} {\bibfnamefont {R.~J.}\ \bibnamefont
  {Birgeneau}}, \bibinfo {author} {\bibfnamefont {M.}~\bibnamefont {Yi}}, \
  and\ \bibinfo {author} {\bibfnamefont {A.}~\bibnamefont {Frano}},\ }\href
  {\doibase 10.1103/PhysRevB.98.035102} {\bibfield  {journal} {\bibinfo
  {journal} {Phys. Rev. B}\ }\textbf {\bibinfo {volume} {98}},\ \bibinfo
  {pages} {035102} (\bibinfo {year} {2018})}\BibitemShut {NoStop}%
\bibitem [{\citenamefont {Liang}\ \emph {et~al.}(2006)\citenamefont {Liang},
  \citenamefont {Bonn},\ and\ \citenamefont {Hardy}}]{Liang2006}%
  \BibitemOpen
  \bibfield  {author} {\bibinfo {author} {\bibfnamefont {R.}~\bibnamefont
  {Liang}}, \bibinfo {author} {\bibfnamefont {D.~A.}\ \bibnamefont {Bonn}}, \
  and\ \bibinfo {author} {\bibfnamefont {W.~N.}\ \bibnamefont {Hardy}},\ }\href
  {\doibase 10.1103/PhysRevB.73.180505} {\bibfield  {journal} {\bibinfo
  {journal} {Phys. Rev. B}\ }\textbf {\bibinfo {volume} {73}},\ \bibinfo
  {pages} {180505} (\bibinfo {year} {2006})}\BibitemShut {NoStop}%
\bibitem [{\citenamefont {Berman}(1976)}]{Berman1976}%
  \BibitemOpen
  \bibfield  {author} {\bibinfo {author} {\bibfnamefont {R.}~\bibnamefont
  {Berman}},\ }\href@noop {} {\emph {\bibinfo {title} {{Thermal conduction in
  solids}}}}\ (\bibinfo  {publisher} {USA: Oxford University Press},\ \bibinfo
  {year} {1976})\BibitemShut {NoStop}%
\bibitem [{\citenamefont {Slack}(1979)}]{Slack1979}%
  \BibitemOpen
  \bibfield  {author} {\bibinfo {author} {\bibfnamefont {G.~A.}\ \bibnamefont
  {Slack}},\ }in\ \href {\doibase
  http://dx.doi.org/10.1016/S0081-1947(08)60359-8} {\emph {\bibinfo {booktitle}
  {Solid State Physics}}},\ \bibinfo {series} {Solid State Physics},
  Vol.~\bibinfo {volume} {34},\ \bibinfo {editor} {edited by\ \bibinfo {editor}
  {\bibfnamefont {H.}~\bibnamefont {Ehrenreich}}, \bibinfo {editor}
  {\bibfnamefont {F.}~\bibnamefont {Seitz}}, \ and\ \bibinfo {editor}
  {\bibfnamefont {D.}~\bibnamefont {Turnbull}}}\ (\bibinfo  {publisher}
  {Academic Press},\ \bibinfo {year} {1979})\ pp.\ \bibinfo {pages} {1 --
  71}\BibitemShut {NoStop}%
\bibitem [{\citenamefont {Slack}(1964)}]{Slack1964}%
  \BibitemOpen
  \bibfield  {author} {\bibinfo {author} {\bibfnamefont {G.~A.}\ \bibnamefont
  {Slack}},\ }\href {\doibase 10.1063/1.1713251} {\bibfield  {journal}
  {\bibinfo  {journal} {Journal of Applied Physics}\ }\textbf {\bibinfo
  {volume} {35}},\ \bibinfo {pages} {3460} (\bibinfo {year}
  {1964})}\BibitemShut {NoStop}%
\bibitem [{\citenamefont {Cahill}\ \emph {et~al.}(1992)\citenamefont {Cahill},
  \citenamefont {Watson},\ and\ \citenamefont {Pohl}}]{Cahill1992}%
  \BibitemOpen
  \bibfield  {author} {\bibinfo {author} {\bibfnamefont {D.~G.}\ \bibnamefont
  {Cahill}}, \bibinfo {author} {\bibfnamefont {S.~K.}\ \bibnamefont {Watson}},
  \ and\ \bibinfo {author} {\bibfnamefont {R.~O.}\ \bibnamefont {Pohl}},\
  }\href {\doibase 10.1103/PhysRevB.46.6131} {\bibfield  {journal} {\bibinfo
  {journal} {Phys. Rev. B}\ }\textbf {\bibinfo {volume} {46}},\ \bibinfo
  {pages} {6131} (\bibinfo {year} {1992})}\BibitemShut {NoStop}%
\bibitem [{\citenamefont {Vandersande}\ and\ \citenamefont
  {Wood}(1986)}]{Vandersande1986}%
  \BibitemOpen
  \bibfield  {author} {\bibinfo {author} {\bibfnamefont {J.~W.}\ \bibnamefont
  {Vandersande}}\ and\ \bibinfo {author} {\bibfnamefont {C.}~\bibnamefont
  {Wood}},\ }\href {\doibase 10.1080/00107518608211003} {\bibfield  {journal}
  {\bibinfo  {journal} {Contemporary Physics}\ }\textbf {\bibinfo {volume}
  {27}},\ \bibinfo {pages} {117} (\bibinfo {year} {1986})}\BibitemShut
  {NoStop}%
\bibitem [{\citenamefont {Langenberg}\ \emph {et~al.}(2016)\citenamefont
  {Langenberg}, \citenamefont {Ferreiro-Vila}, \citenamefont {Leborán},
  \citenamefont {Fumega}, \citenamefont {Pardo},\ and\ \citenamefont
  {Rivadulla}}]{Langenberg2016}%
  \BibitemOpen
  \bibfield  {author} {\bibinfo {author} {\bibfnamefont {E.}~\bibnamefont
  {Langenberg}}, \bibinfo {author} {\bibfnamefont {E.}~\bibnamefont
  {Ferreiro-Vila}}, \bibinfo {author} {\bibfnamefont {V.}~\bibnamefont
  {Leborán}}, \bibinfo {author} {\bibfnamefont {A.~O.}\ \bibnamefont
  {Fumega}}, \bibinfo {author} {\bibfnamefont {V.}~\bibnamefont {Pardo}}, \
  and\ \bibinfo {author} {\bibfnamefont {F.}~\bibnamefont {Rivadulla}},\ }\href
  {\doibase 10.1063/1.4966220} {\bibfield  {journal} {\bibinfo  {journal} {APL
  Materials}\ }\textbf {\bibinfo {volume} {4}},\ \bibinfo {pages} {104815}
  (\bibinfo {year} {2016})}\BibitemShut {NoStop}%
\bibitem [{\citenamefont {Steigmeier}\ and\ \citenamefont
  {Abeles}(1964)}]{Steigmeier1964}%
  \BibitemOpen
  \bibfield  {author} {\bibinfo {author} {\bibfnamefont {E.~F.}\ \bibnamefont
  {Steigmeier}}\ and\ \bibinfo {author} {\bibfnamefont {B.}~\bibnamefont
  {Abeles}},\ }\href {\doibase 10.1103/PhysRev.136.A1149} {\bibfield  {journal}
  {\bibinfo  {journal} {Phys. Rev.}\ }\textbf {\bibinfo {volume} {136}},\
  \bibinfo {pages} {A1149} (\bibinfo {year} {1964})}\BibitemShut {NoStop}%
\bibitem [{\citenamefont {Martelli}\ \emph {et~al.}(2018)\citenamefont
  {Martelli}, \citenamefont {Jim\'enez}, \citenamefont {Continentino},
  \citenamefont {Baggio-Saitovitch},\ and\ \citenamefont
  {Behnia}}]{Martelli2018}%
  \BibitemOpen
  \bibfield  {author} {\bibinfo {author} {\bibfnamefont {V.}~\bibnamefont
  {Martelli}}, \bibinfo {author} {\bibfnamefont {J.~L.}\ \bibnamefont
  {Jim\'enez}}, \bibinfo {author} {\bibfnamefont {M.}~\bibnamefont
  {Continentino}}, \bibinfo {author} {\bibfnamefont {E.}~\bibnamefont
  {Baggio-Saitovitch}}, \ and\ \bibinfo {author} {\bibfnamefont
  {K.}~\bibnamefont {Behnia}},\ }\href {\doibase
  10.1103/PhysRevLett.120.125901} {\bibfield  {journal} {\bibinfo  {journal}
  {Phys. Rev. Lett.}\ }\textbf {\bibinfo {volume} {120}},\ \bibinfo {pages}
  {125901} (\bibinfo {year} {2018})}\BibitemShut {NoStop}%
\bibitem [{Pho()}]{Phonons}%
  \BibitemOpen
  \href@noop {} {}\bibinfo {note} {Here we note that the idea of imposing a
  ``Planckian'' relaxation time bound on the phonons (in addition to the
  electrons) was first introduced by Zhang {\it et al.} \cite{Zhang5378} in the
  thermal diffusivity study of underdoped YBCO system}\BibitemShut {NoStop}%
\bibitem [{\citenamefont {Berggold}\ \emph {et~al.}(2006)\citenamefont
  {Berggold}, \citenamefont {Lorenz}, \citenamefont {Baier}, \citenamefont
  {Kriener}, \citenamefont {Senff}, \citenamefont {Roth}, \citenamefont
  {Severing}, \citenamefont {Hartmann}, \citenamefont {Freimuth}, \citenamefont
  {Barilo},\ and\ \citenamefont {Nakamura}}]{Berggold2006}%
  \BibitemOpen
  \bibfield  {author} {\bibinfo {author} {\bibfnamefont {K.}~\bibnamefont
  {Berggold}}, \bibinfo {author} {\bibfnamefont {T.}~\bibnamefont {Lorenz}},
  \bibinfo {author} {\bibfnamefont {J.}~\bibnamefont {Baier}}, \bibinfo
  {author} {\bibfnamefont {M.}~\bibnamefont {Kriener}}, \bibinfo {author}
  {\bibfnamefont {D.}~\bibnamefont {Senff}}, \bibinfo {author} {\bibfnamefont
  {H.}~\bibnamefont {Roth}}, \bibinfo {author} {\bibfnamefont {A.}~\bibnamefont
  {Severing}}, \bibinfo {author} {\bibfnamefont {H.}~\bibnamefont {Hartmann}},
  \bibinfo {author} {\bibfnamefont {A.}~\bibnamefont {Freimuth}}, \bibinfo
  {author} {\bibfnamefont {S.}~\bibnamefont {Barilo}}, \ and\ \bibinfo {author}
  {\bibfnamefont {F.}~\bibnamefont {Nakamura}},\ }\href {\doibase
  10.1103/PhysRevB.73.104430} {\bibfield  {journal} {\bibinfo  {journal} {Phys.
  Rev. B}\ }\textbf {\bibinfo {volume} {73}},\ \bibinfo {pages} {104430}
  (\bibinfo {year} {2006})}\BibitemShut {NoStop}%
\bibitem [{\citenamefont {Armitage}\ \emph {et~al.}(2010)\citenamefont
  {Armitage}, \citenamefont {Fournier},\ and\ \citenamefont
  {Greene}}]{Armitage2010}%
  \BibitemOpen
  \bibfield  {author} {\bibinfo {author} {\bibfnamefont {N.~P.}\ \bibnamefont
  {Armitage}}, \bibinfo {author} {\bibfnamefont {P.}~\bibnamefont {Fournier}},
  \ and\ \bibinfo {author} {\bibfnamefont {R.~L.}\ \bibnamefont {Greene}},\
  }\href {\doibase 10.1103/RevModPhys.82.2421} {\bibfield  {journal} {\bibinfo
  {journal} {Rev. Mod. Phys.}\ }\textbf {\bibinfo {volume} {82}},\ \bibinfo
  {pages} {2421} (\bibinfo {year} {2010})}\BibitemShut {NoStop}%
\bibitem [{\citenamefont {Vishik}\ \emph {et~al.}(2010)\citenamefont {Vishik},
  \citenamefont {Lee}, \citenamefont {Schmitt}, \citenamefont {Moritz},
  \citenamefont {Sasagawa}, \citenamefont {Uchida}, \citenamefont {Fujita},
  \citenamefont {Ishida}, \citenamefont {Zhang}, \citenamefont {Devereaux},\
  and\ \citenamefont {Shen}}]{Vishik2010}%
  \BibitemOpen
  \bibfield  {author} {\bibinfo {author} {\bibfnamefont {I.~M.}\ \bibnamefont
  {Vishik}}, \bibinfo {author} {\bibfnamefont {W.~S.}\ \bibnamefont {Lee}},
  \bibinfo {author} {\bibfnamefont {F.}~\bibnamefont {Schmitt}}, \bibinfo
  {author} {\bibfnamefont {B.}~\bibnamefont {Moritz}}, \bibinfo {author}
  {\bibfnamefont {T.}~\bibnamefont {Sasagawa}}, \bibinfo {author}
  {\bibfnamefont {S.}~\bibnamefont {Uchida}}, \bibinfo {author} {\bibfnamefont
  {K.}~\bibnamefont {Fujita}}, \bibinfo {author} {\bibfnamefont
  {S.}~\bibnamefont {Ishida}}, \bibinfo {author} {\bibfnamefont
  {C.}~\bibnamefont {Zhang}}, \bibinfo {author} {\bibfnamefont {T.~P.}\
  \bibnamefont {Devereaux}}, \ and\ \bibinfo {author} {\bibfnamefont {Z.~X.}\
  \bibnamefont {Shen}},\ }\href {\doibase 10.1103/PhysRevLett.104.207002}
  {\bibfield  {journal} {\bibinfo  {journal} {Phys. Rev. Lett.}\ }\textbf
  {\bibinfo {volume} {104}},\ \bibinfo {pages} {207002} (\bibinfo {year}
  {2010})}\BibitemShut {NoStop}%
\bibitem [{\citenamefont {Tallon}\ \emph {et~al.}(1995)\citenamefont {Tallon},
  \citenamefont {Bernhard}, \citenamefont {Shaked}, \citenamefont {Hitterman},\
  and\ \citenamefont {Jorgensen}}]{Tallon1995}%
  \BibitemOpen
  \bibfield  {author} {\bibinfo {author} {\bibfnamefont {J.~L.}\ \bibnamefont
  {Tallon}}, \bibinfo {author} {\bibfnamefont {C.}~\bibnamefont {Bernhard}},
  \bibinfo {author} {\bibfnamefont {H.}~\bibnamefont {Shaked}}, \bibinfo
  {author} {\bibfnamefont {R.~L.}\ \bibnamefont {Hitterman}}, \ and\ \bibinfo
  {author} {\bibfnamefont {J.~D.}\ \bibnamefont {Jorgensen}},\ }\href {\doibase
  10.1103/PhysRevB.51.12911} {\bibfield  {journal} {\bibinfo  {journal} {Phys.
  Rev. B}\ }\textbf {\bibinfo {volume} {51}},\ \bibinfo {pages} {12911}
  (\bibinfo {year} {1995})}\BibitemShut {NoStop}%
\bibitem [{\citenamefont {Merzbacher}(1998)}]{Merzbacher}%
  \BibitemOpen
  \bibfield  {author} {\bibinfo {author} {\bibfnamefont {E.}~\bibnamefont
  {Merzbacher}},\ }\href@noop {} {\emph {\bibinfo {title} {Quantum
  Mechanics}}},\ \bibinfo {edition} {3rd}\ ed.\ (\bibinfo  {publisher} {John
  Wiley $\&$ Sons, Inc},\ \bibinfo {address} {New York / Chichester / Weinheim
  Brisbane / Singapore / Toronto},\ \bibinfo {year} {1998})\ Chap.~\bibinfo
  {chapter} {2}\BibitemShut {NoStop}%
\bibitem [{\citenamefont {Helm}\ \emph {et~al.}(2015)\citenamefont {Helm},
  \citenamefont {Kartsovnik}, \citenamefont {Proust}, \citenamefont {Vignolle},
  \citenamefont {Putzke}, \citenamefont {Kampert}, \citenamefont {Sheikin},
  \citenamefont {Choi}, \citenamefont {Brooks}, \citenamefont {Bittner},
  \citenamefont {Biberacher}, \citenamefont {Erb}, \citenamefont {Wosnitza},\
  and\ \citenamefont {Gross}}]{Helm2015}%
  \BibitemOpen
  \bibfield  {author} {\bibinfo {author} {\bibfnamefont {T.}~\bibnamefont
  {Helm}}, \bibinfo {author} {\bibfnamefont {M.~V.}\ \bibnamefont
  {Kartsovnik}}, \bibinfo {author} {\bibfnamefont {C.}~\bibnamefont {Proust}},
  \bibinfo {author} {\bibfnamefont {B.}~\bibnamefont {Vignolle}}, \bibinfo
  {author} {\bibfnamefont {C.}~\bibnamefont {Putzke}}, \bibinfo {author}
  {\bibfnamefont {E.}~\bibnamefont {Kampert}}, \bibinfo {author} {\bibfnamefont
  {I.}~\bibnamefont {Sheikin}}, \bibinfo {author} {\bibfnamefont {E.-S.}\
  \bibnamefont {Choi}}, \bibinfo {author} {\bibfnamefont {J.~S.}\ \bibnamefont
  {Brooks}}, \bibinfo {author} {\bibfnamefont {N.}~\bibnamefont {Bittner}},
  \bibinfo {author} {\bibfnamefont {W.}~\bibnamefont {Biberacher}}, \bibinfo
  {author} {\bibfnamefont {A.}~\bibnamefont {Erb}}, \bibinfo {author}
  {\bibfnamefont {J.}~\bibnamefont {Wosnitza}}, \ and\ \bibinfo {author}
  {\bibfnamefont {R.}~\bibnamefont {Gross}},\ }\href {\doibase
  10.1103/PhysRevB.92.094501} {\bibfield  {journal} {\bibinfo  {journal} {Phys.
  Rev. B}\ }\textbf {\bibinfo {volume} {92}},\ \bibinfo {pages} {094501}
  (\bibinfo {year} {2015})}\BibitemShut {NoStop}%
\bibitem [{\citenamefont {Balci}\ and\ \citenamefont
  {Greene}(2004)}]{Balci2004}%
  \BibitemOpen
  \bibfield  {author} {\bibinfo {author} {\bibfnamefont {H.}~\bibnamefont
  {Balci}}\ and\ \bibinfo {author} {\bibfnamefont {R.~L.}\ \bibnamefont
  {Greene}},\ }\href {\doibase 10.1103/PhysRevLett.93.067001} {\bibfield
  {journal} {\bibinfo  {journal} {Phys. Rev. Lett.}\ }\textbf {\bibinfo
  {volume} {93}},\ \bibinfo {pages} {067001} (\bibinfo {year}
  {2004})}\BibitemShut {NoStop}%
\bibitem [{\citenamefont {Qazilbash}\ \emph {et~al.}(2005)\citenamefont
  {Qazilbash}, \citenamefont {Koitzsch}, \citenamefont {Dennis}, \citenamefont
  {Gozar}, \citenamefont {Balci}, \citenamefont {Kendziora}, \citenamefont
  {Greene},\ and\ \citenamefont {Blumberg}}]{Qazilbash2005}%
  \BibitemOpen
  \bibfield  {author} {\bibinfo {author} {\bibfnamefont {M.~M.}\ \bibnamefont
  {Qazilbash}}, \bibinfo {author} {\bibfnamefont {A.}~\bibnamefont {Koitzsch}},
  \bibinfo {author} {\bibfnamefont {B.~S.}\ \bibnamefont {Dennis}}, \bibinfo
  {author} {\bibfnamefont {A.}~\bibnamefont {Gozar}}, \bibinfo {author}
  {\bibfnamefont {H.}~\bibnamefont {Balci}}, \bibinfo {author} {\bibfnamefont
  {C.~A.}\ \bibnamefont {Kendziora}}, \bibinfo {author} {\bibfnamefont {R.~L.}\
  \bibnamefont {Greene}}, \ and\ \bibinfo {author} {\bibfnamefont
  {G.}~\bibnamefont {Blumberg}},\ }\href {\doibase 10.1103/PhysRevB.72.214510}
  {\bibfield  {journal} {\bibinfo  {journal} {Phys. Rev. B}\ }\textbf {\bibinfo
  {volume} {72}},\ \bibinfo {pages} {214510} (\bibinfo {year}
  {2005})}\BibitemShut {NoStop}%
\bibitem [{\citenamefont {Millis}\ \emph {et~al.}(2005)\citenamefont {Millis},
  \citenamefont {Zimmers}, \citenamefont {Lobo}, \citenamefont {Bontemps},\
  and\ \citenamefont {Homes}}]{Millis2005}%
  \BibitemOpen
  \bibfield  {author} {\bibinfo {author} {\bibfnamefont {A.~J.}\ \bibnamefont
  {Millis}}, \bibinfo {author} {\bibfnamefont {A.}~\bibnamefont {Zimmers}},
  \bibinfo {author} {\bibfnamefont {R.~P. S.~M.}\ \bibnamefont {Lobo}},
  \bibinfo {author} {\bibfnamefont {N.}~\bibnamefont {Bontemps}}, \ and\
  \bibinfo {author} {\bibfnamefont {C.~C.}\ \bibnamefont {Homes}},\ }\href
  {\doibase 10.1103/PhysRevB.72.224517} {\bibfield  {journal} {\bibinfo
  {journal} {Phys. Rev. B}\ }\textbf {\bibinfo {volume} {72}},\ \bibinfo
  {pages} {224517} (\bibinfo {year} {2005})}\BibitemShut {NoStop}%
\bibitem [{\citenamefont {Legros}\ \emph {et~al.}(2019)\citenamefont {Legros},
  \citenamefont {Benhabib}, \citenamefont {Tabis}, \citenamefont
  {Lalibert{\'e}}, \citenamefont {Dion}, \citenamefont {Lizaire}, \citenamefont
  {Vignolle}, \citenamefont {Vignolles}, \citenamefont {Raffy}, \citenamefont
  {Li}, \citenamefont {Auban-Senzier}, \citenamefont {Doiron-Leyraud},
  \citenamefont {Fournier}, \citenamefont {Colson}, \citenamefont {Taillefer},\
  and\ \citenamefont {Proust}}]{Legros2019}%
  \BibitemOpen
  \bibfield  {author} {\bibinfo {author} {\bibfnamefont {A.}~\bibnamefont
  {Legros}}, \bibinfo {author} {\bibfnamefont {S.}~\bibnamefont {Benhabib}},
  \bibinfo {author} {\bibfnamefont {W.}~\bibnamefont {Tabis}}, \bibinfo
  {author} {\bibfnamefont {F.}~\bibnamefont {Lalibert{\'e}}}, \bibinfo {author}
  {\bibfnamefont {M.}~\bibnamefont {Dion}}, \bibinfo {author} {\bibfnamefont
  {M.}~\bibnamefont {Lizaire}}, \bibinfo {author} {\bibfnamefont
  {B.}~\bibnamefont {Vignolle}}, \bibinfo {author} {\bibfnamefont
  {D.}~\bibnamefont {Vignolles}}, \bibinfo {author} {\bibfnamefont
  {H.}~\bibnamefont {Raffy}}, \bibinfo {author} {\bibfnamefont {Z.~Z.}\
  \bibnamefont {Li}}, \bibinfo {author} {\bibfnamefont {P.}~\bibnamefont
  {Auban-Senzier}}, \bibinfo {author} {\bibfnamefont {N.}~\bibnamefont
  {Doiron-Leyraud}}, \bibinfo {author} {\bibfnamefont {P.}~\bibnamefont
  {Fournier}}, \bibinfo {author} {\bibfnamefont {D.}~\bibnamefont {Colson}},
  \bibinfo {author} {\bibfnamefont {L.}~\bibnamefont {Taillefer}}, \ and\
  \bibinfo {author} {\bibfnamefont {C.}~\bibnamefont {Proust}},\ }\href
  {\doibase 10.1038/s41567-018-0334-2} {\bibfield  {journal} {\bibinfo
  {journal} {Nature Physics}\ }\textbf {\bibinfo {volume} {15}},\ \bibinfo
  {pages} {142} (\bibinfo {year} {2019})}\BibitemShut {NoStop}%
\bibitem [{\citenamefont {Padilla}\ \emph {et~al.}(2005)\citenamefont
  {Padilla}, \citenamefont {Lee}, \citenamefont {Dumm}, \citenamefont
  {Blumberg}, \citenamefont {Ono}, \citenamefont {Segawa}, \citenamefont
  {Komiya}, \citenamefont {Ando},\ and\ \citenamefont {Basov}}]{Padilla2005}%
  \BibitemOpen
  \bibfield  {author} {\bibinfo {author} {\bibfnamefont {W.~J.}\ \bibnamefont
  {Padilla}}, \bibinfo {author} {\bibfnamefont {Y.~S.}\ \bibnamefont {Lee}},
  \bibinfo {author} {\bibfnamefont {M.}~\bibnamefont {Dumm}}, \bibinfo {author}
  {\bibfnamefont {G.}~\bibnamefont {Blumberg}}, \bibinfo {author}
  {\bibfnamefont {S.}~\bibnamefont {Ono}}, \bibinfo {author} {\bibfnamefont
  {K.}~\bibnamefont {Segawa}}, \bibinfo {author} {\bibfnamefont
  {S.}~\bibnamefont {Komiya}}, \bibinfo {author} {\bibfnamefont
  {Y.}~\bibnamefont {Ando}}, \ and\ \bibinfo {author} {\bibfnamefont {D.~N.}\
  \bibnamefont {Basov}},\ }\href {\doibase 10.1103/PhysRevB.72.060511}
  {\bibfield  {journal} {\bibinfo  {journal} {Phys. Rev. B}\ }\textbf {\bibinfo
  {volume} {72}},\ \bibinfo {pages} {060511} (\bibinfo {year}
  {2005})}\BibitemShut {NoStop}%
\bibitem [{\citenamefont {Doiron-Leyraud}\ \emph {et~al.}(2015)\citenamefont
  {Doiron-Leyraud}, \citenamefont {Badoux}, \citenamefont {Ren\'ede~Cotret},
  \citenamefont {Lepault}, \citenamefont {LeBoeuf}, \citenamefont
  {Lalibert\'e}, \citenamefont {Hassinger}, \citenamefont {Ramshaw},
  \citenamefont {Bonn}, \citenamefont {Hardy}, \citenamefont {Liang},
  \citenamefont {Park}, \citenamefont {Vignolles}, \citenamefont {Vignolle},
  \citenamefont {Taillefer},\ and\ \citenamefont {Proust}}]{Leyraud2015}%
  \BibitemOpen
  \bibfield  {author} {\bibinfo {author} {\bibfnamefont {N.}~\bibnamefont
  {Doiron-Leyraud}}, \bibinfo {author} {\bibfnamefont {S.}~\bibnamefont
  {Badoux}}, \bibinfo {author} {\bibfnamefont {S.}~\bibnamefont
  {Ren\'ede~Cotret}}, \bibinfo {author} {\bibfnamefont {S.}~\bibnamefont
  {Lepault}}, \bibinfo {author} {\bibfnamefont {D.}~\bibnamefont {LeBoeuf}},
  \bibinfo {author} {\bibfnamefont {F.}~\bibnamefont {Lalibert\'e}}, \bibinfo
  {author} {\bibfnamefont {E.}~\bibnamefont {Hassinger}}, \bibinfo {author}
  {\bibfnamefont {B.~J.}\ \bibnamefont {Ramshaw}}, \bibinfo {author}
  {\bibfnamefont {D.~A.}\ \bibnamefont {Bonn}}, \bibinfo {author}
  {\bibfnamefont {W.~N.}\ \bibnamefont {Hardy}}, \bibinfo {author}
  {\bibfnamefont {R.}~\bibnamefont {Liang}}, \bibinfo {author} {\bibfnamefont
  {J.~H.~.}\ \bibnamefont {Park}}, \bibinfo {author} {\bibfnamefont
  {D.}~\bibnamefont {Vignolles}}, \bibinfo {author} {\bibfnamefont
  {B.}~\bibnamefont {Vignolle}}, \bibinfo {author} {\bibfnamefont
  {L.}~\bibnamefont {Taillefer}}, \ and\ \bibinfo {author} {\bibfnamefont
  {C.}~\bibnamefont {Proust}},\ }\href {\doibase 10.1038/ncomms7034} {\bibfield
   {journal} {\bibinfo  {journal} {Nature Communications}\ }\textbf {\bibinfo
  {volume} {6}},\ \bibinfo {pages} {6034 EP} (\bibinfo {year}
  {2015})}\BibitemShut {NoStop}%
\bibitem [{\citenamefont {Scanderbeg}\ \emph {et~al.}(2016)\citenamefont
  {Scanderbeg}, \citenamefont {Taylor}, \citenamefont {Baumbach}, \citenamefont
  {Paglione},\ and\ \citenamefont {Maple}}]{Scanderbeg2016}%
  \BibitemOpen
  \bibfield  {author} {\bibinfo {author} {\bibfnamefont {D.~J.}\ \bibnamefont
  {Scanderbeg}}, \bibinfo {author} {\bibfnamefont {B.~J.}\ \bibnamefont
  {Taylor}}, \bibinfo {author} {\bibfnamefont {R.~E.}\ \bibnamefont
  {Baumbach}}, \bibinfo {author} {\bibfnamefont {J.}~\bibnamefont {Paglione}},
  \ and\ \bibinfo {author} {\bibfnamefont {M.~B.}\ \bibnamefont {Maple}},\
  }\href {\doibase 10.1088/0953-8984/28/48/485702} {\bibfield  {journal}
  {\bibinfo  {journal} {Journal of Physics Condensed Matter}\ }\textbf
  {\bibinfo {volume} {28}},\ \bibinfo {pages} {485702} (\bibinfo {year}
  {2016})}\BibitemShut {NoStop}%
\bibitem [{\citenamefont {Sarkar}\ \emph {et~al.}(2018)\citenamefont {Sarkar},
  \citenamefont {Greene},\ and\ \citenamefont {Das-Sarma}}]{Sarkar2018}%
  \BibitemOpen
  \bibfield  {author} {\bibinfo {author} {\bibfnamefont {T.}~\bibnamefont
  {Sarkar}}, \bibinfo {author} {\bibfnamefont {R.~L.}\ \bibnamefont {Greene}},
  \ and\ \bibinfo {author} {\bibfnamefont {S.}~\bibnamefont {Das-Sarma}},\
  }\href {https://arxiv.org/abs/1805.08360} {\enquote {\bibinfo {title}
  {{Anomalous normal state resistivity in superconducting
  La$_{2-x}$Ce$_x$CuO$_4$: Fermi liquid or strange metal}},}\ } (\bibinfo
  {year} {2018}),\ \bibinfo {note} {arXiv:1805.08360
  [cond-mat.str-el]}\BibitemShut {NoStop}%
\bibitem [{\citenamefont {Takenaka}\ \emph {et~al.}(2003)\citenamefont
  {Takenaka}, \citenamefont {Nohara}, \citenamefont {Shiozaki},\ and\
  \citenamefont {Sugai}}]{Takenaka2003}%
  \BibitemOpen
  \bibfield  {author} {\bibinfo {author} {\bibfnamefont {K.}~\bibnamefont
  {Takenaka}}, \bibinfo {author} {\bibfnamefont {J.}~\bibnamefont {Nohara}},
  \bibinfo {author} {\bibfnamefont {R.}~\bibnamefont {Shiozaki}}, \ and\
  \bibinfo {author} {\bibfnamefont {S.}~\bibnamefont {Sugai}},\ }\href
  {\doibase 10.1103/PhysRevB.68.134501} {\bibfield  {journal} {\bibinfo
  {journal} {Phys. Rev. B}\ }\textbf {\bibinfo {volume} {68}},\ \bibinfo
  {pages} {134501} (\bibinfo {year} {2003})}\BibitemShut {NoStop}%
\bibitem [{\citenamefont {Homes}\ \emph {et~al.}(2006)\citenamefont {Homes},
  \citenamefont {Lobo}, \citenamefont {Fournier}, \citenamefont {Zimmers},\
  and\ \citenamefont {Greene}}]{Homes2006}%
  \BibitemOpen
  \bibfield  {author} {\bibinfo {author} {\bibfnamefont {C.~C.}\ \bibnamefont
  {Homes}}, \bibinfo {author} {\bibfnamefont {R.~P. S.~M.}\ \bibnamefont
  {Lobo}}, \bibinfo {author} {\bibfnamefont {P.}~\bibnamefont {Fournier}},
  \bibinfo {author} {\bibfnamefont {A.}~\bibnamefont {Zimmers}}, \ and\
  \bibinfo {author} {\bibfnamefont {R.~L.}\ \bibnamefont {Greene}},\ }\href
  {\doibase 10.1103/PhysRevB.74.214515} {\bibfield  {journal} {\bibinfo
  {journal} {Phys. Rev. B}\ }\textbf {\bibinfo {volume} {74}},\ \bibinfo
  {pages} {214515} (\bibinfo {year} {2006})}\BibitemShut {NoStop}%
\bibitem [{\citenamefont {Perepelitsky}\ \emph {et~al.}(2016)\citenamefont
  {Perepelitsky}, \citenamefont {Galatas}, \citenamefont {Mravlje},
  \citenamefont {\ifmmode~\check{Z}\else \v{Z}\fi{}itko}, \citenamefont
  {Khatami}, \citenamefont {Shastry},\ and\ \citenamefont
  {Georges}}]{Perepelitsky2016}%
  \BibitemOpen
  \bibfield  {author} {\bibinfo {author} {\bibfnamefont {E.}~\bibnamefont
  {Perepelitsky}}, \bibinfo {author} {\bibfnamefont {A.}~\bibnamefont
  {Galatas}}, \bibinfo {author} {\bibfnamefont {J.}~\bibnamefont {Mravlje}},
  \bibinfo {author} {\bibfnamefont {R.}~\bibnamefont {\ifmmode~\check{Z}\else
  \v{Z}\fi{}itko}}, \bibinfo {author} {\bibfnamefont {E.}~\bibnamefont
  {Khatami}}, \bibinfo {author} {\bibfnamefont {B.~S.}\ \bibnamefont
  {Shastry}}, \ and\ \bibinfo {author} {\bibfnamefont {A.}~\bibnamefont
  {Georges}},\ }\href {\doibase 10.1103/PhysRevB.94.235115} {\bibfield
  {journal} {\bibinfo  {journal} {Phys. Rev. B}\ }\textbf {\bibinfo {volume}
  {94}},\ \bibinfo {pages} {235115} (\bibinfo {year} {2016})}\BibitemShut
  {NoStop}%
\bibitem [{\citenamefont {Werman}\ and\ \citenamefont
  {Berg}(2016)}]{Werman2016}%
  \BibitemOpen
  \bibfield  {author} {\bibinfo {author} {\bibfnamefont {Y.}~\bibnamefont
  {Werman}}\ and\ \bibinfo {author} {\bibfnamefont {E.}~\bibnamefont {Berg}},\
  }\href {\doibase 10.1103/PhysRevB.93.075109} {\bibfield  {journal} {\bibinfo
  {journal} {Phys. Rev. B}\ }\textbf {\bibinfo {volume} {93}},\ \bibinfo
  {pages} {075109} (\bibinfo {year} {2016})}\BibitemShut {NoStop}%
\bibitem [{\citenamefont {Bruin}\ \emph {et~al.}(2013)\citenamefont {Bruin},
  \citenamefont {Sakai}, \citenamefont {Perry},\ and\ \citenamefont
  {Mackenzie}}]{ScatterSimilar}%
  \BibitemOpen
  \bibfield  {author} {\bibinfo {author} {\bibfnamefont {J.~A.~N.}\
  \bibnamefont {Bruin}}, \bibinfo {author} {\bibfnamefont {H.}~\bibnamefont
  {Sakai}}, \bibinfo {author} {\bibfnamefont {R.~S.}\ \bibnamefont {Perry}}, \
  and\ \bibinfo {author} {\bibfnamefont {A.~P.}\ \bibnamefont {Mackenzie}},\
  }\href {\doibase 10.1126/science.1227612} {\bibfield  {journal} {\bibinfo
  {journal} {Science}\ }\textbf {\bibinfo {volume} {339}},\ \bibinfo {pages}
  {804} (\bibinfo {year} {2013})},\ \bibinfo {note} {see also references
  therein}\BibitemShut {NoStop}%
\bibitem [{\citenamefont {Peng}\ \emph {et~al.}(1991)\citenamefont {Peng},
  \citenamefont {Li},\ and\ \citenamefont {Greene}}]{Peng1991}%
  \BibitemOpen
  \bibfield  {author} {\bibinfo {author} {\bibfnamefont {J.}~\bibnamefont
  {Peng}}, \bibinfo {author} {\bibfnamefont {Z.}~\bibnamefont {Li}}, \ and\
  \bibinfo {author} {\bibfnamefont {R.}~\bibnamefont {Greene}},\ }\href
  {\doibase https://doi.org/10.1016/0921-4534(91)90300-N} {\bibfield  {journal}
  {\bibinfo  {journal} {Physica C: Superconductivity}\ }\textbf {\bibinfo
  {volume} {177}},\ \bibinfo {pages} {79 } (\bibinfo {year}
  {1991})}\BibitemShut {NoStop}%
\bibitem [{\citenamefont {Eisaki}\ \emph {et~al.}(2004)\citenamefont {Eisaki},
  \citenamefont {Kaneko}, \citenamefont {Feng}, \citenamefont {Damascelli},
  \citenamefont {Mang}, \citenamefont {Shen}, \citenamefont {Shen},\ and\
  \citenamefont {Greven}}]{Eisaki2004}%
  \BibitemOpen
  \bibfield  {author} {\bibinfo {author} {\bibfnamefont {H.}~\bibnamefont
  {Eisaki}}, \bibinfo {author} {\bibfnamefont {N.}~\bibnamefont {Kaneko}},
  \bibinfo {author} {\bibfnamefont {D.~L.}\ \bibnamefont {Feng}}, \bibinfo
  {author} {\bibfnamefont {A.}~\bibnamefont {Damascelli}}, \bibinfo {author}
  {\bibfnamefont {P.~K.}\ \bibnamefont {Mang}}, \bibinfo {author}
  {\bibfnamefont {K.~M.}\ \bibnamefont {Shen}}, \bibinfo {author}
  {\bibfnamefont {Z.-X.}\ \bibnamefont {Shen}}, \ and\ \bibinfo {author}
  {\bibfnamefont {M.}~\bibnamefont {Greven}},\ }\href {\doibase
  10.1103/PhysRevB.69.064512} {\bibfield  {journal} {\bibinfo  {journal} {Phys.
  Rev. B}\ }\textbf {\bibinfo {volume} {69}},\ \bibinfo {pages} {064512}
  (\bibinfo {year} {2004})}\BibitemShut {NoStop}%
\bibitem [{\citenamefont {Cohn}\ \emph {et~al.}(1992)\citenamefont {Cohn},
  \citenamefont {Osofsky}, \citenamefont {Peng}, \citenamefont {Li},\ and\
  \citenamefont {Greene}}]{Cohn1992}%
  \BibitemOpen
  \bibfield  {author} {\bibinfo {author} {\bibfnamefont {J.~L.}\ \bibnamefont
  {Cohn}}, \bibinfo {author} {\bibfnamefont {M.~S.}\ \bibnamefont {Osofsky}},
  \bibinfo {author} {\bibfnamefont {J.~L.}\ \bibnamefont {Peng}}, \bibinfo
  {author} {\bibfnamefont {Z.~Y.}\ \bibnamefont {Li}}, \ and\ \bibinfo {author}
  {\bibfnamefont {R.~L.}\ \bibnamefont {Greene}},\ }\href {\doibase
  10.1103/PhysRevB.46.12053} {\bibfield  {journal} {\bibinfo  {journal} {Phys.
  Rev. B}\ }\textbf {\bibinfo {volume} {46}},\ \bibinfo {pages} {12053 }
  (\bibinfo {year} {1992})}\BibitemShut {NoStop}%
\bibitem [{\citenamefont {Denisova}\ \emph {et~al.}(2014)\citenamefont
  {Denisova}, \citenamefont {Chumilina},\ and\ \citenamefont
  {Denisov}}]{Denisova2014}%
  \BibitemOpen
  \bibfield  {author} {\bibinfo {author} {\bibfnamefont {L.~T.}\ \bibnamefont
  {Denisova}}, \bibinfo {author} {\bibfnamefont {L.~G.}\ \bibnamefont
  {Chumilina}}, \ and\ \bibinfo {author} {\bibfnamefont {V.~M.}\ \bibnamefont
  {Denisov}},\ }\href {\doibase 10.1134/S106378341409008X} {\bibfield
  {journal} {\bibinfo  {journal} {Physics of the Solid State}\ }\textbf
  {\bibinfo {volume} {56}},\ \bibinfo {pages} {1928} (\bibinfo {year}
  {2014})}\BibitemShut {NoStop}%
\end{thebibliography}%

%%%%%%%%%%%%%%%%%%%%%%%%%%%%%%%%%

\end{document}